\documentclass[iop,onecolumn]{emulateapj}

\usepackage{amsmath,amssymb}
\usepackage{graphicx}

\def\msun{\,{\rm M}_\odot}
\def\ms{\,{\rm ms}}
\def\km{\,{\rm km}}
\def\mev{\,{\rm MeV}}

\begin{document}


\title{Two-Dimensional Core-Collapse Supernova Models with Multi-Dimensional Transport} 
\author{Joshua C. Dolence\altaffilmark{1}, Adam Burrows\altaffilmark{1}, Weiqun Zhang\altaffilmark{2}}
\altaffiltext{1}{Department of Astrophysical Sciences, Princeton University, Princeton, NJ 08544; 
jdolence@astro.princeton.edu, burrows@astro.princeton.edu}
\altaffiltext{2}{Center for Computational Sciences and Engineering, Lawrence Berkeley National Laboratory, 
Berkeley, CA 94720; weiqunzhang@lbl.gov}

\begin{abstract}
We present new two-dimensional (2D) axisymmetric neutrino radiation/hydrodynamic models of core-collapse supernova (CCSN) cores.  We use the CASTRO code, which incorporates truly multi-dimensional, multi-group, flux-limited diffusion (MGFLD) neutrino transport, including all relevant $\mathcal{O}(v/c)$ terms.  Our main motivation for carrying out this study is to compare with recent 2D models produced by other groups who have obtained explosions for some progenitor stars and with recent 2D VULCAN results that did not incorporate $\mathcal{O}(v/c)$ terms.  We follow the evolution of 12, 15, 20, and 25 solar-mass progenitors to approximately 600 milliseconds after bounce and do not obtain an explosion in any of these models.  Though the reason for the qualitative disagreement among the groups engaged in CCSN modeling remains unclear, we speculate that the simplifying ``ray-by-ray' approach employed by all other groups may be compromising their results.  We show that ``ray-by-ray' calculations greatly exaggerate the angular and temporal variations of the neutrino fluxes, which we argue are better captured by our multi-dimensional MGFLD approach.  On the other hand, our 2D models also make approximations, making it difficult to draw definitive conclusions concerning the root of the differences between groups.  We discuss some of the diagnostics often employed in the analyses of CCSN simulations and highlight the intimate relationship between the various explosion conditions that have been proposed.  Finally, we explore the ingredients that may be missing in current calculations that may be important in reproducing the properties of the average CCSNe, should the delayed neutrino-heating mechanism be the correct mechanism of explosion.
\end{abstract}

\keywords{(stars:)supernovae: general}

\section{Introduction}
The mechanism underlying the explosive deaths of massive stars remains poorly understood.  \citet{Colg66} first 
suggested that neutrino energy deposition plays a central role in powering core-collapse supernovae (CCSNe).  
Ever since, much of the effort in CCSN theory has focused on building increasingly sophisticated 
neutrino radiation hydrodynamical models, with the hopes of reproducing the properties of 
CCSNe, including the kinetic energies, debris morphologies, nucleosynthetic yields, and the remnant mass, 
spin, and velocity distributions.  Despite this effort, the best models still fall short of accounting 
for any of these properties, much less all of them simultaneously.  Perhaps even more alarming, 
the various groups involved in CCSN modeling often reach qualitatively different conclusions with 
calculations that are ostensibly quite similar, vis-\'{a}-vis whether an explosion even occurs.  

In \citet{Mull12,Mull12_sasi}, results are reported from 2D axisymmetric modeling with the \textsc{Vertex-CoCoNuT} code, 
which uses a conformally-flat spacetime approximation of general relativity \citep{Mull10}.  They find 
explosions for $8.1$-$\msun$, $11.2$-$\msun$, $15$-$\msun$, and $27$-$\msun$ progenitors, but, when 
reported, the explosion energies are $\sim$10 times smaller than the canonical $10^{51}\, {\rm erg}$ energy 
of typical CCSNe.  Similar findings were presented for a variety of other progenitors (Janka et al., Nuclear Astrophysics Workshop, Ringberg Castle, 2014).  Recently, \citet{Hank13} reported results from 
a three-dimensional simulation with the \textsc{Prometheus-Vertex} code of the same $27$-$\msun$ progenitor 
considered in \citet{Mull12_sasi} and found no explosion.  This negative result has been recapitulated for all
other 3D simulations performed recently by this group \citep{Tamb14}, despite their having seen explosions in the corresponding 2D simulations.    

\citet{Suwa10} reports an explosion of a $13$-$\msun$ progenitor in a 2D simulation and \citet{Taki12} finds explosions of an $11.2$-$\msun$ progenitor in both 2D and 3D.  These models neglected the heavy lepton neutrinos, which were recently incorporated with an approximate leakage scheme \citep{Taki13}.  In all cases, the $\nu_e$ and $\bar{\nu}_e$ transport was computed using the isotropic diffusion source approximation (IDSA) \citep{Lieb09}, a crude approximation meant to enable multi-D simulations at minimal cost.  While interesting, their results are difficult to interpret in the context of the viability of the neutrino mechanism, as the authors acknowledge.

Meanwhile, \citet{Brue13} report results of 
2D axisymmetric modeling with their \textsc{Chimera} code.  They consider $12$-$\msun$, $15$-$\msun$, 
$20$-$\msun$, and $25$-$\msun$ progenitors from \citet{Woos07} and find explosions in all cases, 
curiously at almost the same post-bounce time.  They also report energies that are somewhat larger than 
the those reported in \citet{Mull12}, but that still fall short of the $10^{51}\, {\rm erg}$ mark.  Janka at el. (Nuclear Astrophysics Workshop, Ringberg Castle, 2014) recently reported 2D models of the same four progenitors, and found significantly different results, with, for example, the $12$-$\msun$ model not yet exploding more than $700\ms$ after bounce.

Importantly, all of the studies discussed above relied on the so-called 
``ray-by-ray-plus'' approximation of neutrino transport, which replaces the real transport problem 
with a series of independent spherically-symmetric transport solves.  This is a crude approximation 
that introduces large variations in angle and time in the neutrino fluxes and the associated neutrino energy deposition 
so crucial for the neutrino-driven mechanism.  This simplification has yet to be clearly justified, 
and may be producing qualitatively incorrect results, particularly in 2D.  

The only calculations ever performed 
which allow for multidimensional transport were the VULCAN/2D results reported in \citet{Burr06}, \citet{Burr07}, \citet{Ott08}, and \citet{Bran11}, 
and none of these calculations showed a revival of the stalled shock in 2D by the delayed-neutrino mechanism.
The calculations of \citet{Ott08} and \citet{Bran11} were multi-angle as well.
However, these calculations were performed without $\mathcal{O}(v/c)$ transport effects \citep{Hube07}.
We are, therefore, motivated in this paper to perform new 2D multi-group radiation hydrodynamics calculations with
a new code with both multi-D transport (avoiding the simplifications of the ray-by-ray approach) and 
the velocity-dependent terms to determine whether these earlier results were artifacts of the neglect 
of $\mathcal{O}(v/c)$ terms, and for comparison with the 2D results of other groups.
To accomplish this, we have developed the CASTRO radiation hydrodynamics code. CASTRO contains a 
multi-group flux-limited neutrino transport solver, is time-dependent and multidimensional, 
and treats three neutrino species ($\nu_e$, $\bar{\nu}_e$, $\nu_x$, where the $\nu_x$ species 
includes the $\mu$ and $\tau$ neutrinos and their antiparticles), including all relevant $\mathcal{O}(v/c)$ terms.
We find that none of our new 2D calculations, employing the same progenitors as \citet{Brue13}, explode by the delayed-neutrino mechanism.
With this paper, we describe our results and speculate on the reasons for the different outcomes we find.
Since all other groups are using the ray-by-ray approach, we suggest that one reason for the different outcomes
may be in the handling of multi-D transport.  In 2D, the axial sloshing motions, often not seen in 3D \citep{Burr12}, may be reinforcing the errors in the ray-by-ray approach
and leading to a qualitatively incorrect outcome.  In 3D, these axial sloshing effects are often absent, and the ray-by-ray approach
may be less anomalous (due to the greater sphericity of the hydrodynamics), so the lack
of explosions seen by the Garching group in 3D, when they observe explosions for the same progenitors in 2D, remains
puzzling.


\section{Numerics and Setup}

We use the CASTRO code to carry out our CCSN simulations \citep{Almg10,Zhan11,Zhan13}.  
CASTRO is a second-order, Eulerian, compressible, Godunov-type, radiation hydrodynamics 
code that uses block-structured adaptive mesh refinement to simultaneously refine 
in both space and time.  Simulations can be performed in 1D (spherical), 2D (cylindrical), and 3D (Cartesian).  The hydrodynamic 
updates use piecewise-parabolic reconstruction with higher-order limiters to preserve 
accuracy at smooth extrema, an approximate Riemann solver detailed in \citet{Almg10}, and incorporates 
full corner coupling in the directionally unsplit integration.  In this work, we make 
use of the multi-group flux-limited diffusion (MGFLD) neutrino transport solver detailed 
in \citet{Zhan13}.  

Currently, multi-dimensional multi-angle
transport is not feasible and will require the exascale.
Our comoving frame multi-group flux-limited diffusion (MGFLD) formulation includes $\mathcal{O}(v/c)$ terms that can
result (correctly) in significant differences in the dynamic diffusion limit where
radiation transport is dominated by motion of the fluid \citep{Cast04}.
The approach used in CASTRO
splits the system into three parts: 1) a part that couples the radiation
and fluid in a hyperbolic subsystem with a piecewise parabolic
method (PPM) with characteristic tracing and full corner
coupling \citep{Mill02}, 2) another part that is a system of coupled
parabolic equations that evolves radiation diffusion over all the
groups (along with $Y_e$ in the neutrino case) and source-sink
terms, and 3) a final part that performs frequency-space advection.
The hyperbolic subsystem is solved explicitly with a high-order
Godunov scheme as part of the hydrodynamic component of the algorithm,
whereas the parabolic part is solved implicitly with a first-order
backward Euler method.  Frequency-space advection is performed using
a standard approach based on the method of lines.  The frequency-space advection has its
own CFL condition for stability and, if necessary, subcycling in time is employed in order to satisfy the
frequency-space CFL condition. The primary computational expense of
the radiation is in the solution of linear systems as part of the
iteration over energy groups.  We rely on the {\sl hypre} library
\citep{Falg02} for solving these systems on large parallel machines.

CASTRO uses a hybrid parallelization strategy based on MPI + OpenMP
using the BoxLib framework \citep{Almg10}. The basic strategy
is to distribute grids within the AMR hierarchy to computational nodes.
This provides a natural coarse-grained approach to distributing the
computational work.  A dynamic load balancing technique is needed to
adjust the load.  Although the code supports both a heuristic knapsack
algorithm and a space-filling curve algorithm for load balancing, the
data-locality properties make the space-filling curve the method of
choice for problems with radiation. The main advantages of the CASTRO code
are the efficiency due to the use of AMR, the temporal sub-cycling, and
the accuracy due to the coupling of radiation force into the Riemann solver.

CASTRO incorporates 1) a complicated set of opacity tables for the
various neutrino species that includes weak-magnetism,
ion-ion correlation, and ion form-factor corrections,
2) various extant nuclear equations of state (including the
Shen [default] and various Lattimer and Swesty equations of state), 3) inelastic
neutrino-electron and neutrino-nucleon scattering, and 4) a
temporal sub-cycling algorithm that accelerates computation
by many factors over traditional codes.  CASTRO
treats three neutrino species ($\nu_e$, $\bar{\nu}_e$, $\nu_x$, where 
the $\nu_x$ species includes the $\mu$ and $\tau$ neutrinos and their antiparticles).  In this work, 
for computational expediency, we neglect in the 2D simulations energy-group coupling processes such as 
inelastic scattering, and justify this choice in Section~\ref{1D} through the comparison of model shock radius 
evolutions in 1D, with and without them.  We also adopt the Shen equation of state \citep{Shen98a,Shen98b}. 

We follow the collapse, bounce, and subsequent evolution of four progenitors in 1D and 2D.  The progenitors are nonrotating, solar metallicity models with zero-age main sequence masses of $12$-$\msun$, $15$-$\msun$, $20$-$\msun$, and $25$-$\msun$ \citep{Woos07}.  These are the same progenitors recently considered by \citet{Brue13} and Janka et al. (Nuclear Astrophysics Workshop, Ringberg Castle, 2014).  Our numerical grid has $0.5\km$ resolution in the inner $\sim$$128\km$, $1\km$ resolution out to $\sim$$270\km$, never worse than $4\km$ resolution anywhere beneath the shock, and extends out to $5120\km$.

\section{Results of Two-Dimensional Simulations}
\label{results}

\subsection{Global Structure}
\label{global}
\begin{figure}
\includegraphics[width=\textwidth]{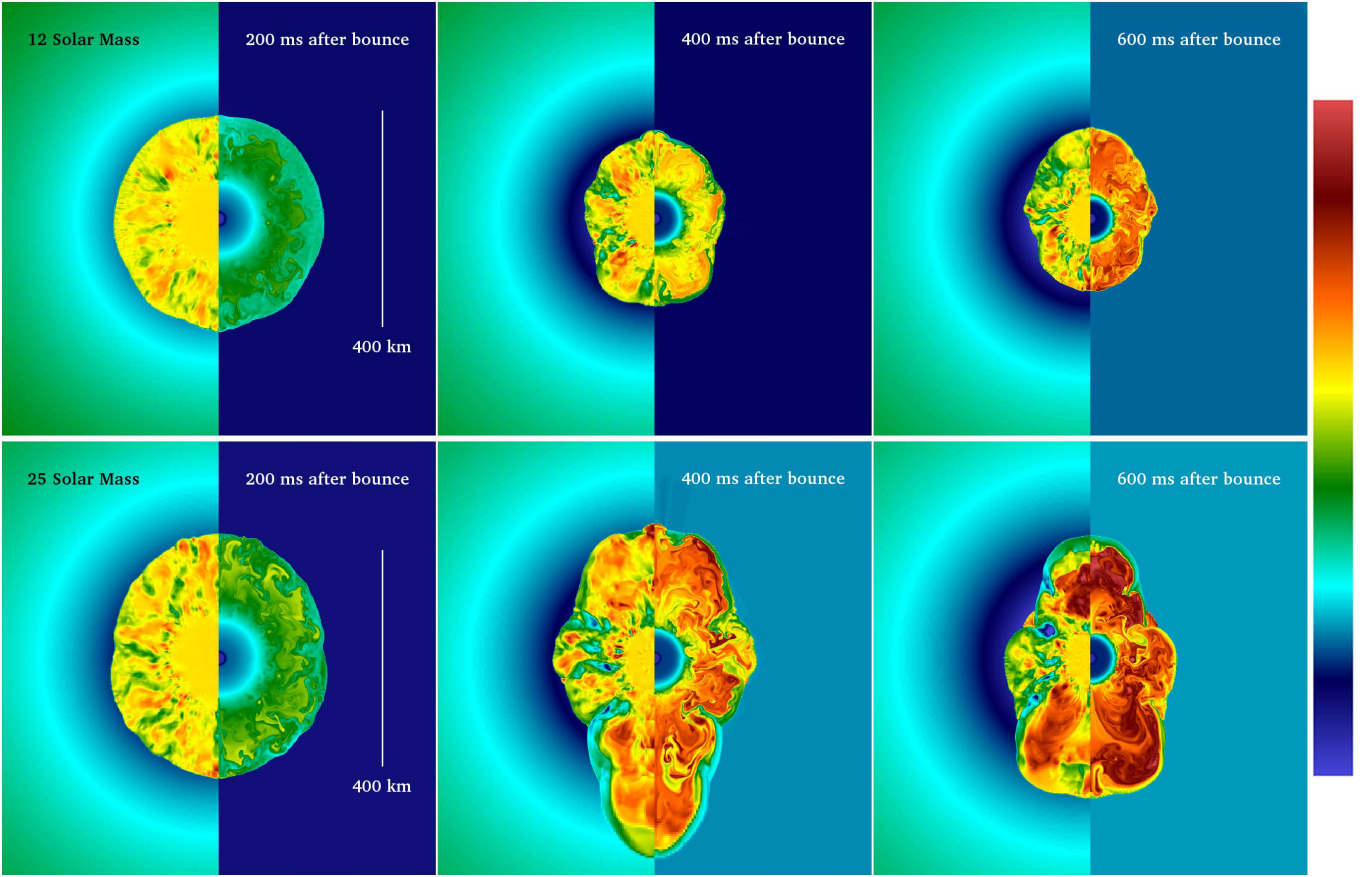}
\caption{Snapshots of the radial velocity (left half of each image) and entropy (right half of each image) for the $12$-$\msun$ (top three images) and $25$-$\msun$ (bottom three images) models at 200, 400, and $600\ms$ after bounce.  The colorbar, shown at right, ranges from $[-6\times 10^4,4\times 10^4]\,{\rm km}\,{\rm s}^{-1}$ for the radial velocity and from $[0,25]\,k_B\,{\rm baryon}^{-1}$ for the entropy.  Convective activity can be clearly seen as relatively low entropy accretion streams and upwelling higher entropy plumes.}
\label{fig:snapshots}
\end{figure}

Figure~\ref{fig:snapshots} shows representative snapshots of the radial velocity and entropy from our $12$-$\msun$ and $25$-$\msun$ models.  At $200\ms$ after bounce, both models are nearly spherical and show weak convective activity with a characteristic angular scale corresponding to $\ell\sim10$ ($\ell$ is spherical harmonic degree) as is clearly seen in the radial velocities.  By $400\ms$ after bounce, both models have become more aspherical, indicating stronger convective activity that is at larger angular scales (smaller $\ell$).  The snapshots at $600\ms$ after bounce look qualitatively similar to those at $400\ms$ after bounce, showing large scale convective motions including relatively low entropy accretion streams and higher entropy buoyant plumes.

\begin{figure}
\includegraphics[width=0.8\textwidth]{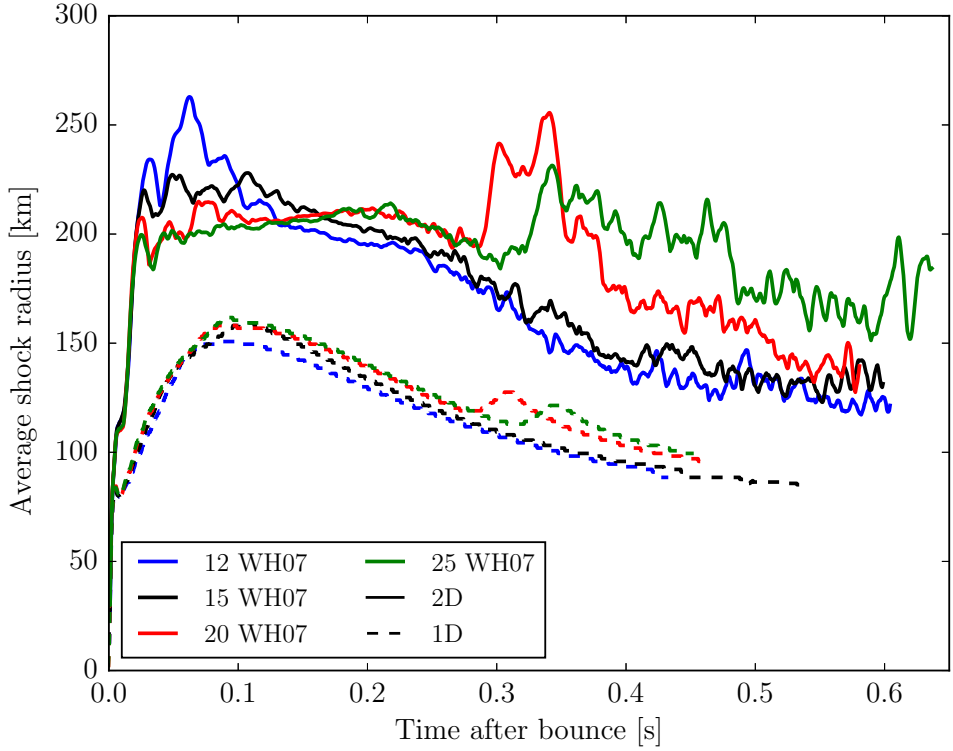}
\caption{Evolutions of the average shock radii for our 1D and 2D simulations.  The 2D models diverge from their 1D counterparts almost immediately, likely due to a brief but vigorous phase of prompt convection seeded by our aspherical grid.  The 2D average shock radii then remain larger throughout the evolution.  Around $300\ms$ after bounce, the $20$-$\msun$ and $25$-$\msun$ models show jumps due to the accretion of the Si/O interface and the jump is fractionally larger in 2D than in 1D.}
\label{fig:rshock}
\end{figure}

Figure~\ref{fig:rshock} shows the evolutions of the average shock radii for our 1D and 2D models.  Not surprisingly, our 1D models do not explode, consistent with nearly all spherically symmetric CCSN calculations.  In 2D, despite evolving the models to $\gtrsim$$600\ms$ after bounce, we do not find explosions.  Nevertheless, there are interesting differences between 1D and 2D models.  Perhaps most importantly, the 2D stalled shock radii are consistently larger by a factor 1.3--2, as found in essentially all 1D-2D comparisons \citep[e.g.][]{Mull12}.  This is often attributed to aspherical instabilities, namely convection and the standing accretion shock instability (SASI), both increasing the neutrino heating efficiency in the gain region and providing turbulent pressure support \citep[e.g.][]{Hera92,Burr95,Jank96}.  In our models, we see convective activity develop immediately as the bounce shock moves out.  This brief phase of ``prompt convection,'' seeded by perturbations introduced by our aspherical grid, drives the shock to large radii much faster than in the corresponding 1D models.  By $\sim$$20\ms$ after bounce, all four 2D models have shock radii of $\sim$$200\km$ and these shocks stall $\sim$$60\ms$ earlier than the 1D models, which never reach radii much beyond $\sim$$150\km$.  Between 300 and 400$\ms$, the $20$-$\msun$ and $25$-$\msun$ models show jumps in the average shock radii, both in 1D and 2D.  These jumps correspond to the accretion of the Si/O interface in these models, wherein $\dot{M}$ drops suddenly and the shock responds by moving outward.  In 1-D, the jump is only about $10\%$ of the pre-interface shock radius, but the effect is 2--3 times larger in 2D.  Apparently and interestingly, the sensitivity to the accretion of shelves increases with shock radius. 

\begin{figure}
\includegraphics[width=0.8\textwidth]{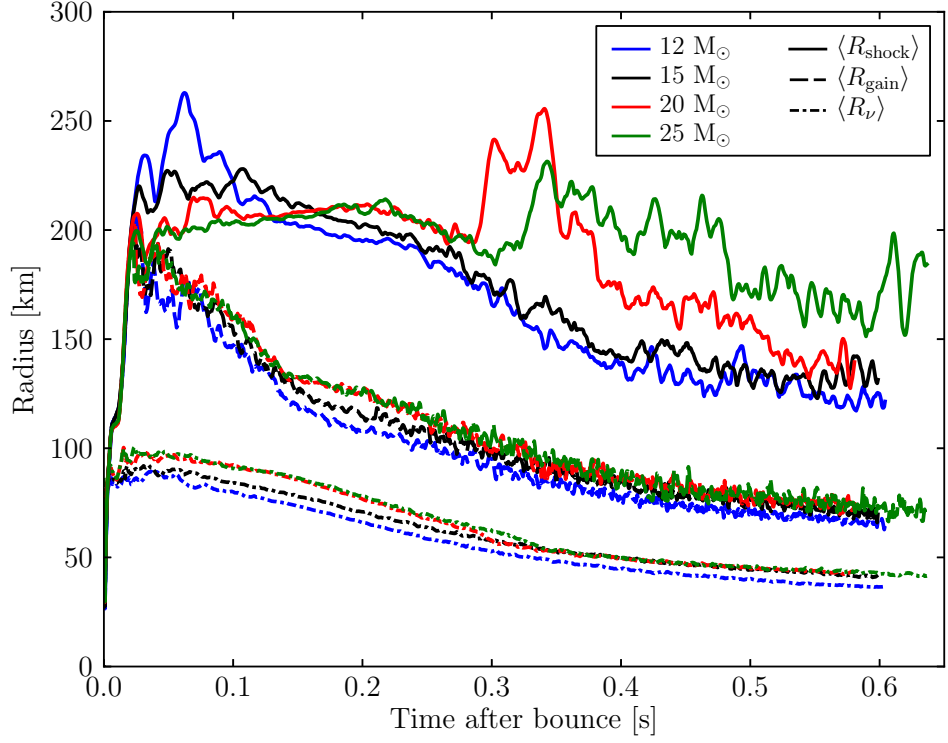}
\caption{Evolutions of the average shock, gain, and inner radii, as defined in Sec.~\ref{global}.  After the initial expansions of the bounce shocks, the gain regions form and the gain radii move inward.  This is accompanied by the inward migration of the inner cooling radius.  Though the shock radii respond to the accretion of the Si/O interface in the $20$-$\msun$ and $25$-$\msun$ models, the gain and inner cooling radius do not.}
\label{fig:radii}
\end{figure}

The shock radii are only one aspect of the global structure that determines if and when explosions commence.  The gain radius, $R_g(\theta)$, is defined as the radius above which the net rate of energy deposition in the gas, integrated over all neutrino energies and summed over species, is positive.  The gain region, bounded by the gain and shock radii, contains the mass where neutrino energy deposition, occurring at a rate
\begin{equation}\label{eq:heating_rate}
(\mathcal{H} - \mathcal{C})_{\rm gain} = \sum_s \int d\Omega \int_{R_g(\theta)}^{R_s(\theta)} r^2 dr \int_{\varepsilon_{s}^{\rm min}}^{\varepsilon_{s}^{\rm max}} (c\kappa_{s,\varepsilon} E_{s,\varepsilon} - j_{s,\varepsilon}) d\varepsilon\,,
\end{equation}
 is thought to lead to shock revival.  In Eq.~\ref{eq:heating_rate}, $s\in\left\{\nu_e,\bar{\nu}_e,\nu_x\right\}$, $R_s(\theta)$ is the angle-dependent shock radius, $\varepsilon$ is neutrino energy, $\kappa$ is the absorption cross section, $E$ is the neutrino energy density, and $j$ is the volume emissivity.  As important is the cooling region where accreted material cools and settles onto the proto-neutron star, undermining the pressure support crucial to reviving the stalled shock.  We define a net cooling rate analogous to the heating rate above as
\begin{equation}
(\mathcal{H} - \mathcal{C})_{\rm cool} = \sum_s \int d\Omega \int_{R_{\nu_s}(\theta)}^{R_g(\theta)} r^2 dr \int_{\varepsilon_{\nu_s}^{\rm min}}^{\varepsilon_{\nu_s}^{\rm max}} (c\kappa_{s,\varepsilon} E_{s,\varepsilon} - j_{s,\varepsilon}) d\varepsilon\,,
\end{equation}
where $R_{\nu_s}$ is the angle-, species-, and energy-dependent neutrinosphere radius defined implicitly by
\begin{equation}
\int_{R_{\nu_s}}^\infty \kappa dr = \frac{2}{3}\,.
\end{equation}
This definition roughly accounts for all the optically-thin cooling in the region, but does not include diffusive flux from beneath the $\tau=2/3$ surface.  We define an effective inner cooling radius as
\begin{equation}
R_{\rm in} = \frac{\sum_s \sum_\varepsilon |w_s| R_{\nu_s}}{\sum_s \sum_\varepsilon |w_s|}\,,
\end{equation}
where
\begin{equation}
w_s = \int_{R_{\nu_s}}^\infty (c\kappa E - j) r^2 dr
\end{equation}
is an integral of the net neutrino energy deposition along a column (at fixed $\theta$) and should be understood to be angle-, species-, and energy-dependent.  This inner cooling radius is related to the neutrinosphere radii of the dominant cooling agents.  The region between $R_{\rm in}$ and $R_g$ is cooling rapidly by optically-thin neutrino emission, predominantly through the $\nu_e$ and $\bar{\nu}_e$ species, and represents at least one component of the accretion luminosity.  Figure~\ref{fig:radii} shows the evolutions of the shock, gain, and inner radii for our four 2D models.  After a short-lived initial transient, the gain and inner radii generally recede in all four models, reflecting the contraction of the inner proto-neutron star.

\subsection{Diagnostic Quantities}
\label{diagnostics}
\begin{figure}
\includegraphics[width=0.8\textwidth]{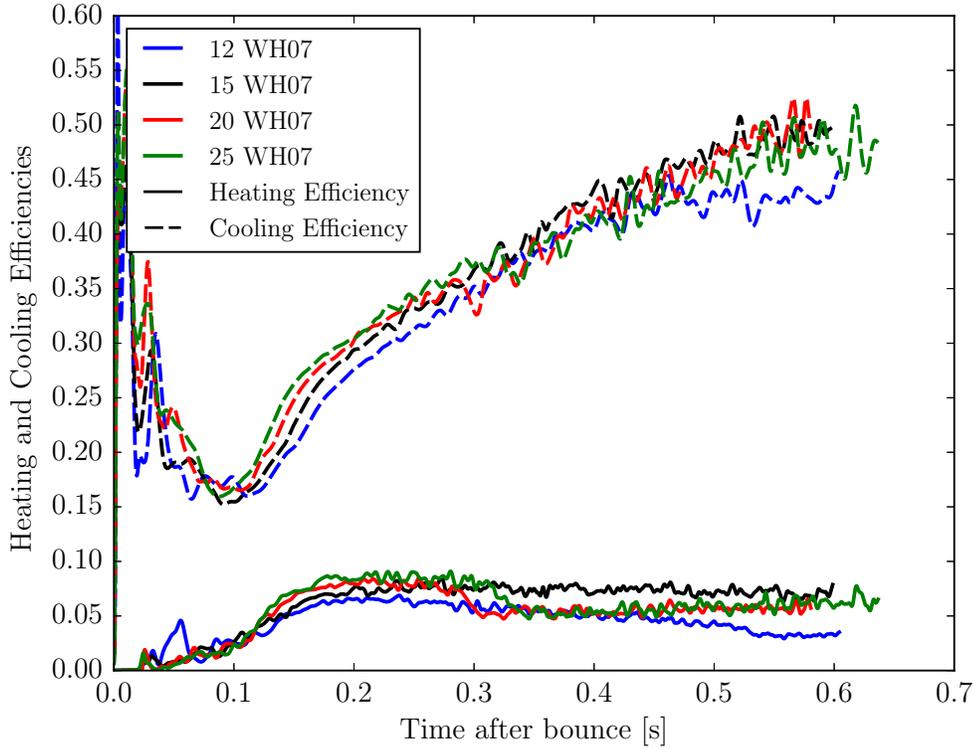}
\caption{Evolutions of the heating and cooling efficiencies as defined in Sec.~\ref{diagnostics}.  The heating efficiencies rise slowly into a broad peak around $200\ms$ in all models, whereas the cooling efficiencies show minima around $100\ms$ before growing for the remainder of the simulations.}
\label{fig:efficiencies}
\end{figure}

The evolutions of certain diagnostic quantities have proven useful in distilling insight from the complicated multi-D dynamics of pre-explosion supernova cores.  The heating efficiency, $\epsilon_h$, is defined as the ratio of the net heating in the gain region to the sum $L_{\nu_e}+L_{\bar{\nu}_e}$.  We also define an analogous cooling efficiency, $\epsilon_c\equiv-(\mathcal{H}-\mathcal{C})_{\rm cool}/(L_{\nu_e}+L_{\bar{\nu}_e})$.  Both efficiencies are shown in Fig.~\ref{fig:efficiencies}.  The heating efficiencies initially rise slowly and show broad peaks around $200\ms$ after bounce.  The peak values range from 6\% to 9\%, monotonically increasing with progenitor mass.  At low optical depths, the heating efficiency is equivalent to an effective optical depth, giving a heating rate $(L_{\nu_e}+L_{\bar{\nu}_e}) (1-\exp{(-\tau_{\rm eff})})\approx (L_{\nu_e}+L_{\bar{\nu}_e}) \tau_{\rm eff}$.  The basic character of the heating efficiencies can be understood by a simple estimate of $\tau_{\rm eff}$.  First, write $\tau_{\rm eff}\sim\langle \rho\rangle_g \sigma \Delta R$, where $\langle \rho\rangle_g$ is the average density in the gain region, $\sigma$ is a characteristic neutrino absorption cross section per unit mass, and $\Delta R=\langle R_s\rangle - \langle R_g\rangle$.  The average density is easily computed from the gain mass and shock and gain radii.  We write the cross section as $\sigma\approx\sigma_0 (\varepsilon/\varepsilon_0)^2\sim\sigma_0 (R_0/R_{\rm in})^2$, where the 0-subscript signifies some characteristic value.  This last scaling is justified by the approximately inverse relationship between root-mean-square neutrino energy and inner cooling radius.  Finally, we arrive at
\begin{equation}\label{eq:tau}
\tau_{\rm eff}\sim \frac{3\sigma_0 M_g}{4\pi} \frac{R_s-R_g}{R_s^3-R_g^3} \left(\frac{R_0}{R_{\rm in}}\right)^2\,.
\end{equation}
This estimate reproduces the heating efficiencies reasonably well, accounting, for example, for the slow decline in the $12$-$\msun$ model, the nearly constant value for the $15$-$\msun$ model, and the rise in the latter half of the $20$-$\msun$ and $25$-$\msun$ models.  The virtue of this estimate is that it shows how the heating efficiencies depend on the global structure of the solutions.

The cooling efficiency $\epsilon_c$ represents the fraction of the $\nu_e$ and $\bar{\nu}_e$ luminosities arising from optically thin cooling.  Since some of the accreted material advects into optically thick regions before cooling appreciably, the net cooling rate $(\mathcal{H}-\mathcal{C})_{\rm cool}$ entering into the definition of $\epsilon_c$ is not equal to the total accretion luminosity, which must be $\approx GM_{\rm pns}\dot{M}_{\rm pns}/R_{\rm pns}$ (pns$\equiv$proto-neutron star), at least in an integral-averaged sense.  Rather, $\epsilon_c$ represents a response function to sudden changes in $\dot{M}$.  For example, if the accretion rate drops due to the accretion of a composition interface, the immediate response in the luminosity (delayed somewhat by the advection time between the shock and cooling region) will be to drop fractionally by approximately $\epsilon_c \dot{M}/\dot{M}_0$, where $\dot{M}_0$ and $\dot{M}$ are the pre- and post-interface accretion rates, respectively.  The noticeable shelves accreted in the $12$-$\msun$, $20$-$\msun$, and $25$-$\msun$ models confirm this basic picture.  Apparently, a low cooling efficiency may make a model prone to explosion when shelves are accreted since, though $\dot{M}$ can drop appreciably, the luminosity responds only weakly, leaving the model closer to (or perhaps beyond) the critical luminosity for explosion \citep{Burr13}.  Evidently, this effect is not sufficient to drive explosions in our models.

\begin{figure}
\includegraphics[width=0.9\textwidth]{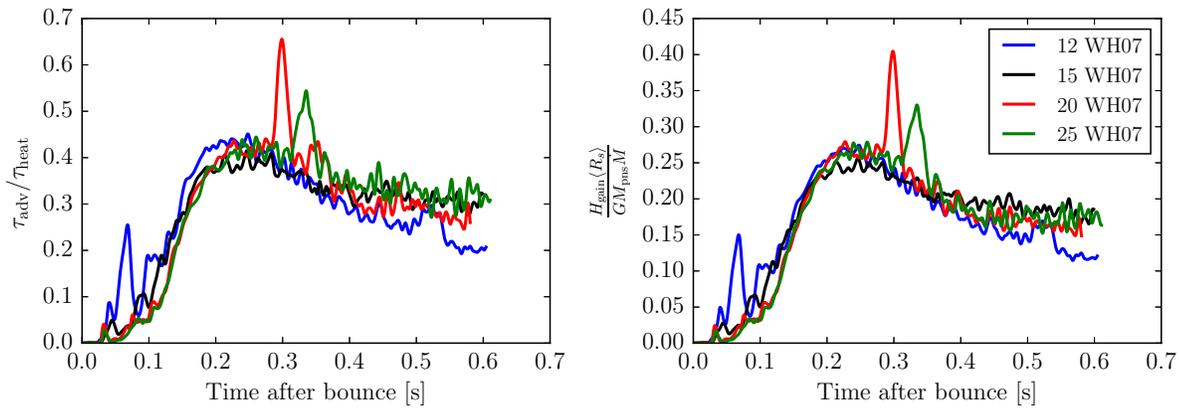}
\caption{The left panel shows the evolutions of the ratio of advection to heating time scales.  All four models show a broad peak around $200\ms$ after bounce, around the same time the heating efficiencies are largest.  The narrow peaks around $300\ms$ for the $20$-$\msun$ and $25$-$\msun$ models correspond to the accretion of the Si/O interface when $\dot{M}$ drops by a factor of two.  The right panel shows the evolutions of the dimensionless ratios of heating to accretion power.  Aside from the overall normalization, these curves have a striking resemblance to those shown in the left panel, demonstrating that the ratio of advection to heating timescales is intimately related to the ratio of heating to accretion power.}
\label{fig:crit_panel}
\end{figure}

A variety of ``explosion conditions'' have been proposed in the literature.  The most popular considers the ratio of the timescale for advection through the gain region to the heating timescale.  Define the advection timescale as $\tau_{\rm adv} = M_{\rm gain}/\dot{M}$ and the heating timescale as
\begin{equation}
\tau_{\rm heat} = \frac{\int_{\rm gain} (\rho e - \rho e_0) dV}{(\mathcal{H}-\mathcal{C})_{\rm gain}}\,,
\end{equation}
where $e_0$ is approximately the zero-point of the EOS, i.e. $e_0=e(\rho,Y_e,T_{\rm min})$ where $T_{\rm min}$ is the lowest temperature of the tabulated EOS ($10^{-2}$ MeV).  The left panel of Fig.~\ref{fig:crit_panel} shows the evolutions of this ratio for all four models.  Surprisingly, the curves all show the same basic structure, peaking around $200\ms$ after bounce before slowly declining.  Superimposed on these slowly varying curves, the $20$-$\msun$ and $25$-$\msun$ models show narrow peaks around $300\ms$ after bounce, corresponding to the accretion of the Si/O interfaces in these models when $\dot{M}$ drops abruptly by a factor of two.

Another interesting dimensionless quantity that arises naturally in discussing critical conditions for explosion \citep[see, e.g.,][]{Dole13} is the ratio of heating in the gain region to the accretion power $\mathcal{H}_g R_s/(G M \dot{M})$, shown in the right panel of Fig.~\ref{fig:crit_panel}.
Comparing the left and right panels of Fig.~\ref{fig:crit_panel}, it is clear that these dimensionless ratios have very similar behaviors.  Writing the heating in the gain region $\mathcal{H}_g = h_g M_g$, where $h_g$ is the specific heating rate and $M_g$ is the mass in the gain region, and noting that the zero-point-subtracted specific internal energy is of order $G M/R_s$, it is easy to see that $\mathcal{H}_g R_s/(G M \dot{M})\sim \tau_{\rm adv}/\tau_{\rm heat}$, so the similarities between the left and right panels of Fig.~\ref{fig:crit_panel} are not surprising.  Similarly, the ``antesonic'' condition, proposed by \citet{Pejc12}, is easily related to these other conditions.  Apparently, $\tau_{\rm adv}/\tau_{\rm heat} \sim \mathcal{H}_g R_s/(G M \dot{M}) \propto c_s^2/v_{\rm esc}^2$.

While all three ratios are interesting diagnostics and can clearly indicate when explosions are underway, none seem to have a well-determined critical threshold above which explosions are inevitable.  In particular, the value of one is not necessarily associated with transitioning to explosion.  The critical values likely depend in detail on how these quantities are defined, and may be below or above one, but reasonable definitions seem to at least give critical values of order one.  Of course, since we do not obtain explosions, we are unable to determine these critical values for our models and definitions.

\section{Comparisons with One-Dimensional Results}
\label{1D}

Multidimensional effects seem to be crucial if the neutrino mechanism succeeds in producing explosions.  A variety of effects may be important.  The post-shock turbulent flow, driven by convective and/or SASI activity, leads to longer dwell times in the gain region on average, exposing the accreted material to more net heating \citep{Murp08,Dole13}.  The turbulent flow itself gives rise to Reynolds stresses that can help support the shock \citep{Murp13}.  Convection in the proto-neutron star can advect neutrinos closer to their neutrinosphere, shortening their escape time and boosting the diffusive luminosity from the core \citep{Dess06,Mull14}.  Even the efficiency of cooling through the accretion luminosity can vary between dimensions.  We have already explored the former two effects in parametrized setups \citep{Dole13,Murp13}, so here we focus on characterizing how the neutrino luminosities and average energies differ between 1D and 2D models.

\begin{figure}
\includegraphics[width=\textwidth]{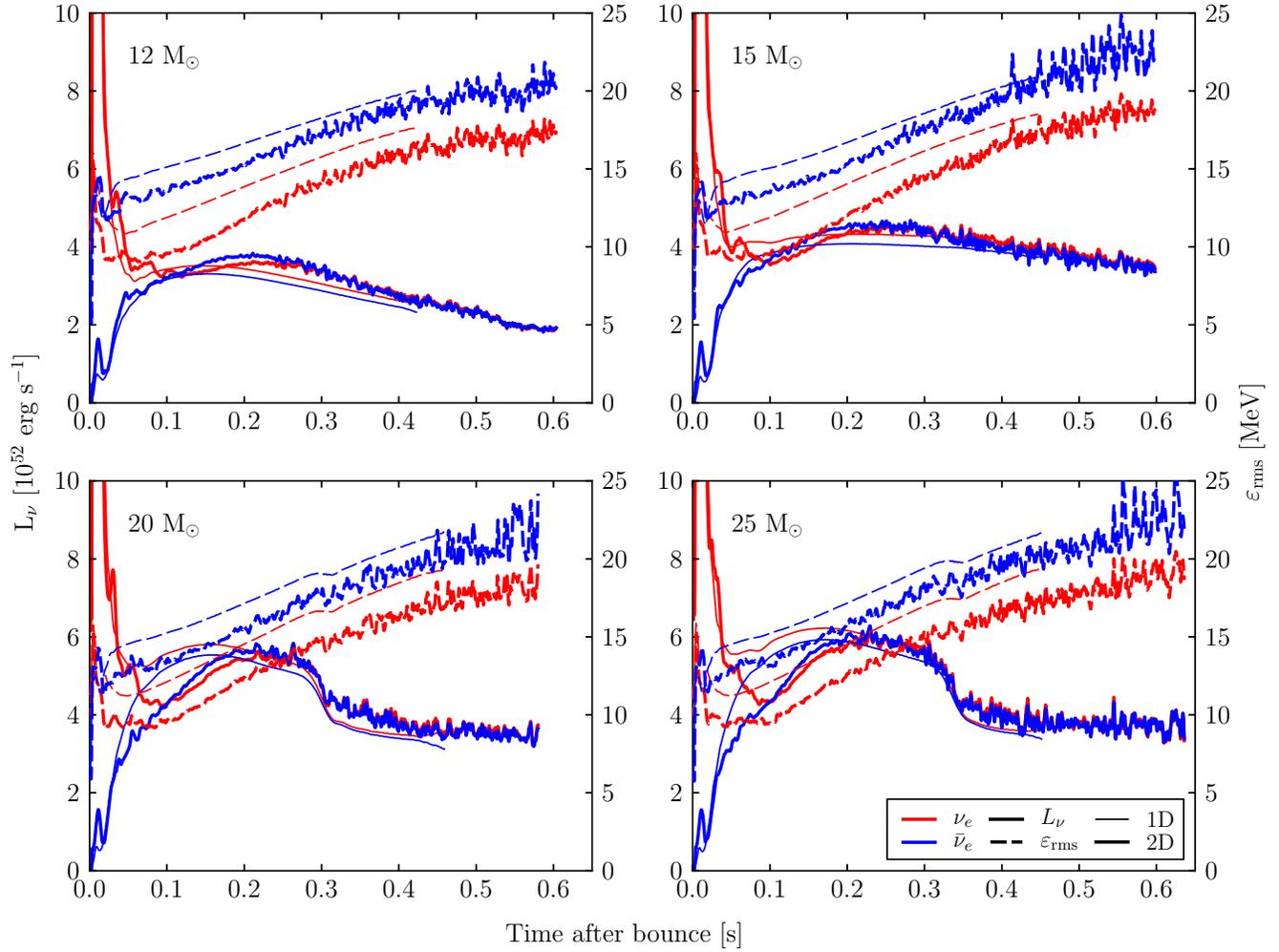}
\caption{Evolution of the neutrino luminosities (solid lines, left ordinate) and root-mean-square neutrino energies (dashed lines, right ordinate) for our 1D (thin lines) and 2D (thick lines) simulations of all four progenitors.  The 1D models tend to have slightly higher rms neutrino energies, but, at least at later times, slightly lower luminosities.}
\label{fig:le_1d2d}
\end{figure}

Figure~\ref{fig:le_1d2d} shows the evolutions of the $\nu_e$ and $\bar{\nu}_e$ luminosities and root-mean-square neutrino energies, as measured in the laboratory frame at $500\km$ and defined by
\begin{equation}
\varepsilon_{\rm rms} = \sqrt{\frac{\int_{\varepsilon_{\rm min}}^{\varepsilon_{\rm max}} \varepsilon^2 F_\varepsilon d\varepsilon}{\int_{\varepsilon_{\rm min}}^{\varepsilon_{\rm max}} F_\varepsilon d\varepsilon}}\, ,
\end{equation}
for all four progenitor models.  Some differences between 1D and 2D seem to be model-independent.  For example, the rms energies are systematically higher in 1D than in 2D by $\sim$5--10\%, consistent with results obtained by other groups \citep{Bura06}.  The luminosities, on the other hand, do not show such a generic difference.  Around $100\ms$ post-bounce, the $20$-$\msun$ and $25$-$\msun$ models show deficits of up to $\sim$$20\%$ in $\nu_e$ and $\bar{\nu}_e$ luminosities in 2D relative to 1D.  Beyond $\sim$$200\ms$ post bounce, all the 2D models tend to have higher $\nu_e$ and $\bar{\nu}_e$ luminosities by $\sim$$5\%$, likely an effect of Ledoux convection in the proto-neutron star, driven by the destabilizing lepton gradient \citep{Bura06,Dess06}.

The idea of a critical luminosity for explosion at a given mass accretion rate was first discussed in \citet{Burr93}.  \citet{Murp08} later showed, in the context of parametrized modeling, that the critical luminosity is lower in 2D than in 1D, a result confirmed by other studies \citep{Hank13,Dole13,Couc13_2d3d}.  Unfortunately, the critical curves that emerged from these studies are not easily adapted to self-consistent radiation hydrodynamic models.  One reason is that the critical luminosity almost certainly depends on quantities other than $L_{\nu_e}$ and $\dot{M}$.  For example, it may depend on the proto-neutron star mass and radius and/or the shock radius, as suggested in the discussion of the dimensionless quantities in Section~\ref{diagnostics}.  It also depends on the structure of the flow beneath the gain layer, which can be quite different between parametrized and self-consistent models.  Nevertheless, the lower critical luminosity in 2D very likely remains in self-consistent models.

\begin{figure}
\includegraphics[width=0.8\textwidth]{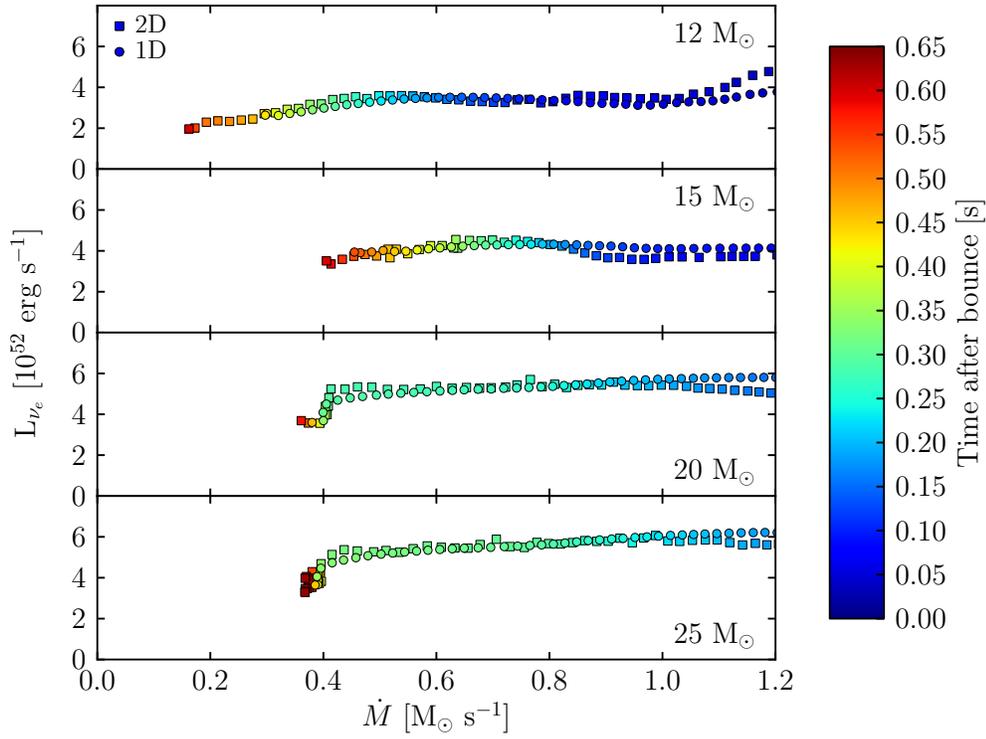}
\caption{Trajectories in the $L_{\nu_e}$-$\dot{M}$ plane for all four progenitors in 1D and 2D.  Interestingly, all four models show a roughly horizontal track, indicating that the luminosities are roughly constant while the accretion rates drop.  The late time luminosity drops of the $20$-$\msun$ and $25$-$\msun$ models are somewhat delayed relative to the drops in the accretion rates associated with the Si/O interfaces.  Evidently, since none of our models explode, the critical curve must be to the left of the data in each plot.}
\label{fig:crit}
\end{figure}

In Fig.~\ref{fig:crit}, we show the evolutions in the $L_{\nu_e}$-$\dot{M}$ plane for all four progenitors in both 1D and 2D.  All of the models show relatively flat curves until late times, indicating that the neutrino luminosities remain roughly constant while the accretion rate drops.  It is during this phase that a crossing of critical luminosity curve (and subsequent explosion) seem most likely, though this does not occur in our models.  As in Fig.~\ref{fig:le_1d2d}, the 2D models tend to show higher $\nu_e$ luminosities at late times relative to 1D.  Naively, this might suggest that 2D models are easier to explode not only because the critical curve is lower, but also because 2D models tend to have higher luminosities at a given mass accretion rate.  Of course, this argument neglects other important differences between 1D and 2D including the higher rms neutrino energies in 1D, for example.  Rewriting Eq.~\ref{eq:tau} for the effective optical depth (or, equivalently, the heating efficiency) with $M_g=\dot{M} \tau_{\rm adv}$ and $R_0/R_{\rm in}=\varepsilon/\varepsilon_0$, we arrive at
\begin{equation}
\frac{\tau_{\rm eff}^{\rm 2D}}{\tau_{\rm eff}^{\rm 1D}} \sim  \left(\frac{\tau_{\rm adv}^{\rm 2D}}{\tau_{\rm adv}^{\rm 1D}}\right) \left[ \left(\frac{R_s-R_g}{R_s^3-R_g^3}\right)^{\rm 2D} \left(\frac{R_s^3-R_g^3}{R_s-R_g}\right)^{\rm 1D} \right] \left(\frac{\varepsilon_{\rm rms}^{\rm 2D}}{\varepsilon_{\rm rms}^{\rm 1D}}\right)^2\;.
\end{equation}
The first term on the right hand side is the ratio of the advection timescales, which favors 2D.  The second term (in square brackets) favors 1D since the gain volume is typically much larger in 2D.  As we show in Fig.~\ref{fig:le_1d2d}, the last term given by the square of the ratio of rms neutrino energies favors 1D.  Evidently, though some effects in 2D tend toward lower heating efficiencies, these are overwhelmed by the different advection timescales, leading to higher heating efficiencies in 2D.  Now consider the diagnostic given by the ratio of heating in the gain region to accretion power, which Fig.~\ref{fig:crit_panel} shows is intimately related to the ratio of advection to heating timescales.  It is easy to show that the 2D to 1D ratio of this diagnostic is
\begin{equation}
\left(\frac{H R_s}{G M \dot{M}}\right)^{\rm 2D}\left/\left(\frac{H R_s}{G M \dot{M}}\right)^{\rm 1D}\right. \approx \left(\frac{L_{\nu_e}^{\rm 2D}}{L_{\nu_e}^{\rm 1D}}\right) \left(\frac{\tau_{\rm eff}^{\rm 2D}}{\tau_{\rm eff}^{\rm 1D}}\right) \left(\frac{R_s^{\rm 2D}}{R_s^{\rm 1D}}\right)\;.
\end{equation}
At least after the first 100--200$\ms$, all of these terms favor 2D, making 2D models much easier to explode than corresponding 1D models.

\begin{figure}
\includegraphics[width=0.8\textwidth]{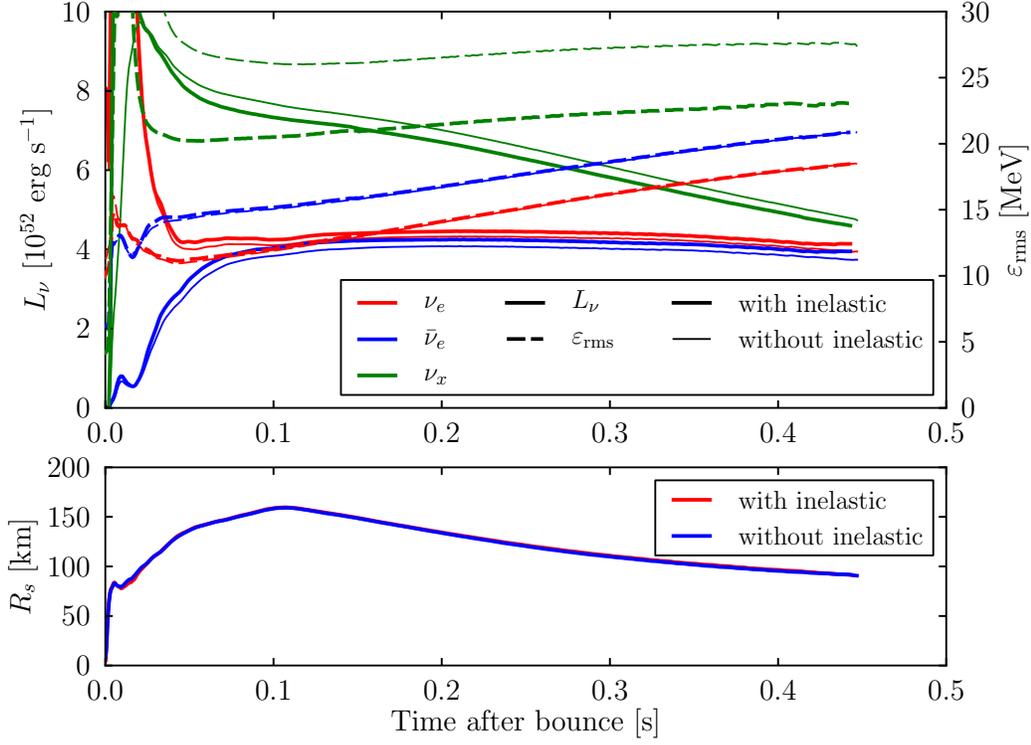}
\caption{Comparison of 1D simulations of the $15$-$\msun$ progenitor, with and without inelastic scattering on electrons.  The largest change is in the rms energy of the $\nu_x$ species, which is lowered by $\sim$$5\mev$ with inelastic scattering included.  The shock radii, shown in the bottom panel, are almost identical, demonstrating that the effects of inelastic scattering on electrons is quite small in terms of changes in the global structure of the flow.}
\label{fig:inelastic}
\end{figure}

In the 1D and 2D models discussed thus far, we have ignored inelastic scattering processes.  Figure~\ref{fig:inelastic} shows results from two 1D simulations of the $15$-$\msun$ progenitor that differ only in whether inelastic scattering on electrons is included.  The most marked difference is in the rms energy of the $\nu_x$ species, which is lower with inelastic scattering included by about $5\mev$.  In principle, depositing $5\mev$ per $\nu_x$ neutrino represents a significant source of heating, but this energy deposition occurs mainly at large optical depths for the $\nu_e$ and $\bar{\nu}_e$ species, so the net effect is substantially muted.  For example, the $\nu_e$ and $\bar{\nu}_e$ luminosities are higher by only 3--4\% and the rms neutrino energies are higher by only $\sim$$0.5\%$ in the model with inelastic scattering.  The $\nu_x$ luminosity with inelastic scattering adjusts so that the total neutrino luminosity, summed over all species, is nearly identical to the model without inelastic scattering, with fractional differences typically less than $0.1\%$.  Thus, though inelastic scattering on electrons does lead to small changes in the luminosities and average energies of each species, the net effect on the structure as a whole is quite small.  For example, the shock radii shown in Fig.~\ref{fig:inelastic} are almost identical, differing by at most $2\%$ and typically less than $1\%$.

\section{Angular and Temporal Variations in the Neutrino Sector}
\label{angle}

\subsection{A Criticism of ``Ray-by-ray''}
One of the motivations for carrying out this study is to compare our MGFLD neutrino transport results with those obtained with the \textit{de facto} standard ``ray-by-ray'' formulation currently used by all other groups carrying out radiation hydrodynamic simulations of core-collapse supernovae.  \citet{Ott08} carried out both MGFLD and full multi-angle transport calculations and found that, though MGFLD is only approximate in the semi-transparent and free streaming regimes, the two methods actually agree quite well in the supernova context.  No such comparison with ``ray-by-ray'' has yet been discussed \citep[but see][for a comparison of multi-angle and ``ray-by-ray'' transport results based on time-independent snapshots of multidimensional simulations]{Sumi14}.  Here, we take a small step towards this comparison, which captures one of our main criticisms of the technique --- the unphysically high degree of angular and temporal correlation between the matter and neutrino radiation fields.

An argument often made in the core-collapse community is that the spatio-temporal variations in the flow effectively average out the error made in employing the ``ray-by-ray'' approach \citep{Bura06a,Mezz14}.  However, both \citet{Bura06a} and \citet{Mezz14} acknowledge that such averaging can only be approximate and that artifacts may remain in ``ray-by-ray'' calculations.  For example, \citet{Bura06a}, in an attempt to address this concern, post-processed their ``ray-by-ray'' models by recomputing the heating in the gain region using an angularly averaged radiation field.  When integrated over the volume of the gain region and also time-averaged, they found good agreement between the ``ray-by-ray'' and angularly-averaged ``ray-by-ray'' heating rates \citep[see also][]{Sumi14}.  Beyond this zeroth order comparison, however, larger differences begin to emerge.  For example, \citet{Bura06a} find that the heating of downflows and high-entropy bubbles, when analyzed separately, show differences larger than 10\% in the time-averaged heating rates as computed by ``ray-by-ray'' and angularly-averaged calculations.  This finding strongly reinforces our concern about the artificially high degree of correlation between matter and neutrino sectors.  Importantly, the most crucial question with regards to ``ray-by-ray,'' how it effects the time-dependent hydrodynamic response of the system, remains open.  We do not attempt to address that here and therefore make no claim that we have proven that ``ray-by-ray'' calculations are wholly wrong.  Instead, we aim only to show some ways in which ``ray-by-ray'' calculations fall short, and to suggest that these shortcomings may make the current crop of supernova modeling efforts in the community difficult to reconcile.

\begin{figure}
\includegraphics[width=0.5\textwidth]{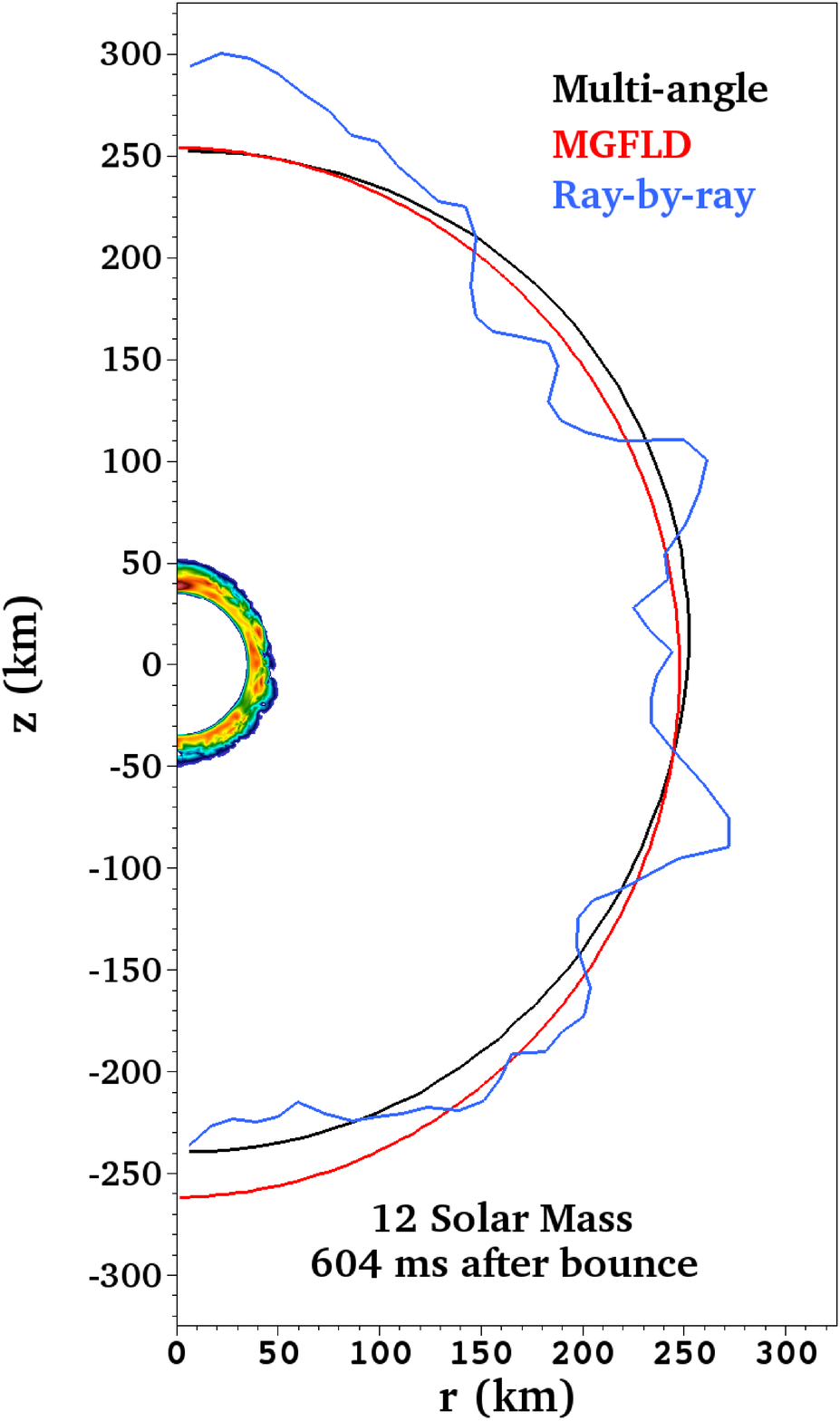}
\caption{Comparison of the transport schemes as discussed in Sec.~\ref{angle}.  The pseudocolor map shows $\rho T^6 \exp(-\rho/10^{11})$, roughly representative of the local cooling rate.  The lines show the normalized $\nu_e$-fluxes measured in the laboratory frame at $250\km$.  The normalization scales the fluxes so that their solid-angle average is 250 on the plot.  Angular variations in the hydrodynamic sector are clearly, and unphysically, represented in the $\nu_e$-fluxes for the ``ray-by-ray'' scheme, while the MGFLD and multi-angle fluxes show much smoother variations characteristic of the integral nature of transport.}
\label{fig:rbr_comp}
\end{figure}

To that end, we consider a snapshot from our 2D simulation of the $12$-$\msun$ model $604\ms$ after bounce (i.e., at the end of the run).  We extract the energy integrated fluxes of the electron neutrinos at $250\km$ as computed by CASTRO's MGFLD solver.  With this snapshot, we do two more transport calculations, ignoring velocity- and time-dependence and scattering.  The first calculation gives the full angle- and energy-dependent specific intensities at discrete latitudes at $250\km$.  We then integrate over energy and take the first moment to get the fluxes.  Finally, we compute the angle- and energy-dependent specific intensities under the ``ray-by-ray'' assumption, then integrate over energy and take the first moment to recover the ``ray-by-ray'' fluxes.  We describe our techniques for the multi-angle and ``ray-by-ray'' calculations in Appendix~\ref{appendix}.  The simulation snapshot and the results of these transport calculations are shown in Fig.~\ref{fig:rbr_comp}.  The MGFLD and full multi-angle transport results agree quite well in the general character of the angular variation, despite the neglect of velocity- and time-dependence and of scattering in the simplified multi-angle calculation.  The variation is smooth and has a small amplitude.  Though the low-$\ell$ modes have similar amplitudes compared with the other two schemes, the ``ray-by-ray'' results show much more intermediate- to high-$\ell$ power, manifest as wild variations that close inspection reveals are highly correlated with structures in the flow.  The peak-to-peak variation in the ``ray-by-ray'' result is about a factor of five larger than for MGFLD and about a factor of four larger than for multi-angle transport.  While this may already be a concern, the most serious failing of the ``ray-by-ray'' approach is that these large variations are tightly correlated with hydrodynamic structures along each ray.  The neutrino radiation field is produced by an integral over many sources at depth, an effect which can only be truly captured by doing full multi-angle transport and seems reasonably well approximated by MGFLD, but which a ``ray-by-ray'' approach can never reliably reproduce.  The ``ray-by-ray'' flux depends only on the profile along a given radial ray, introducing large variations which should be washed away by the integral character of transport and which are highly correlated with the hydrodynamic variations of the flow.  We note that our choice of measuring the fluxes at $250\km$ was motivated by the desire to include transport through the entire gain region, but choosing a radius far from the neutrinosphere tends to emphasize the differences between the schemes.  However, one must measure the fluxes very close to the neutrinosphere for the methods to show comparable fluctuation, and even then the ``ray-by-ray'' scheme exaggerates the variation, though to a lesser degree than when viewed at $250\km$.

\subsection{Angular Variations}
\label{angular_variations}
\begin{figure}
\includegraphics[width=0.8\textwidth]{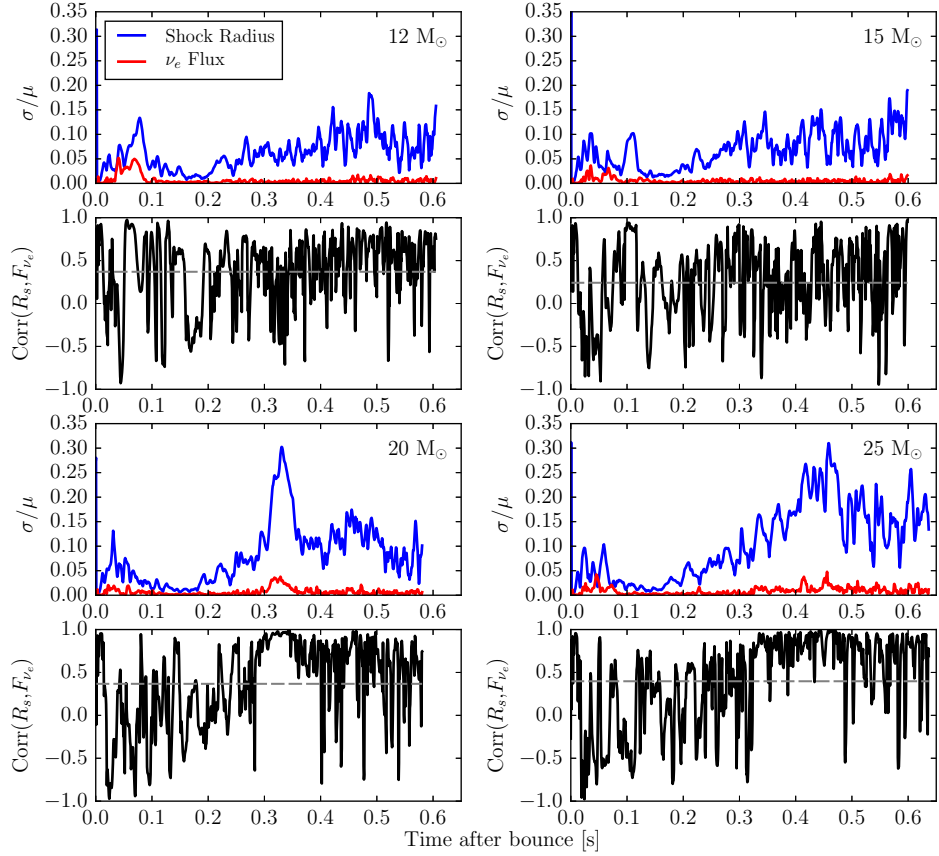}
\caption{For all four progenitor models, the top panel shows the standard deviations of the shock radius and energy integrated $\nu_e$ flux at $500\km$, normalized by their respective means.  The shock radius shows a fractional variation about an order-of-magnitude larger than the $\nu_e$-fluxes.  The bottom panels show the cross-correlations of these quantities, as defined in Sec.~\ref{angular_variations}.  The dashed gray lines show the value of the time-averaged cross-correlations. See text in Section \ref{angular_variations} for a discussion.}
\label{fig:shock_lum_corr}
\end{figure}

As has been discussed previously, the angular variations of the neutrino radiation field are muted relative to variations in the matter sector \citep{Bran11}.  In Fig.~\ref{fig:shock_lum_corr}, we show the normalized standard deviations of the shock radii and $\nu_e$-fluxes (at $500\km$) as a function of time.  We define the standard deviation ($\sigma$) normalized by the mean ($\mu$) of quantity $Q$ as 
\begin{equation}
\frac{\sigma}{\mu} = \frac{1}{\langle Q\rangle}\left(\frac{\int_0^\pi (Q(\theta)-\langle Q(\theta)\rangle)^2 \sin\theta d\theta}{\int_0^\pi \sin\theta d\theta}\right)^{1/2}\;,
\end{equation}
where the angle brackets indicate solid-angle averaging.  Typically, the fractional angular variations in the shock radii are about an order-of-magnitude larger than the corresponding variations in the fluxes.  We also show the cross-correlations of the shock radii and $\nu_e$-fluxes, defined by
\begin{equation}
Corr(R_s,F_{\nu_e}) = \frac{1}{2}\int_0^\pi \frac{(R_s - \langle R_s \rangle)(F_{\nu_e}-\langle F_{\nu_e}\rangle)}{\sigma_{R_s} \sigma_{F_{\nu_e}}} \sin\theta d\theta\;,
\end{equation}
where $\sigma$ is the standard deviation, computed as above, but without the normalization by the mean.  All the models show a positive correlation on average, as indicated by the gray dashed lines in the figure.  The $20$-$\msun$ and $25$-$\msun$ models show a period $\sim$$100\ms$ long of consistently high correlation.  Interestingly, these phases follow immediately after the accretion of the significant Si/O interfaces in these models, where the accretion rates drop by a factor $\approx$2.  Importantly, however, though the shock and $\nu_e$-fluxes are correlated in all models, the amplitude of the $\nu_e$-flux variation is much smaller than one would find with a ``ray-by-ray'' calculation, so the degree to which these variations couple may be quite different than in ``ray-by-ray'' calculations \citep{Burr13}.

\begin{figure}
\includegraphics[width=0.8\textwidth]{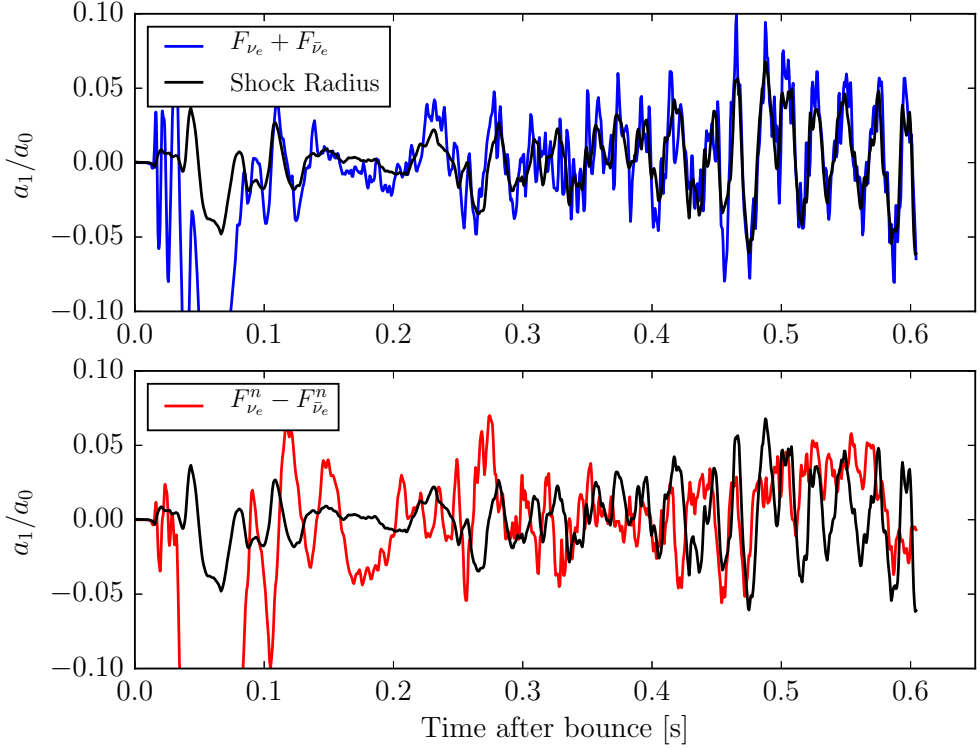}
\caption{The bottom panel shows the evolutions of the normalized dipolar component $a_1/a_0$ of the lepton-number flux $F^n_{\nu_e}-F^n_{\bar{\nu}_e}$ and shock radius, the former multiplied by five for ease of comparison.  The signals appear only weakly correlated, which is confirmed quantitatively by computing their cross-correlation.  The amplitude of the lepton-number flux asymmetry is also quite small, about an order of magnitude smaller than reported in \citet{Tamb14}.  Taken together, these results seem at odds with the expectations of the LESA model described in \citet{Tamb14}.  The top panel shows the evolutions of the normalized dipolar component of $F_{\nu_e}+F_{\bar{\nu}_e}$ along with the shock, the former multiplied by ten.  These quantities are clearly much more correlated than the quantities in the bottom panel.}
\label{fig:lesa}
\end{figure}

\citet{Tamb14} recently reported systematic asymmetries in the net lepton-number emission in several 3D models and dubbed this feature LESA (Lepton-number Emission Self-sustained Asymmetry).  To investigate whether we see a similar phenomenon, we decompose the lepton-number flux $F^n_{\nu_e} - F^n_{\bar{\nu}_e}$ of our $12$-$\msun$ model ($n$ indicates number instead of energy flux) into spherical harmonics and focus on the evolution of the normalized dipolar component $a_1/a_0$.  In the model described in \citet{Tamb14}, deviations from spherical symmetry in the lepton-number flux are intimately related to corresponding deviations in the shock structure.  In the bottom panel of Fig.~\ref{fig:lesa}, we show the evolutions of $a_1/a_0$ for the lepton-number flux and the shock surface.  For ease of comparison, we have multiplied $a_1/a_0$ for the lepton-number flux by five.  We see no sign of a strong correlation between the two quantities.  Indeed, the normalized cross-correlation as a function of time lag shows a broad feature around zero offset, with a modest peak value of $\sim$0.3 (1 indicates perfectly correlated signals) at $-21\ms$, comparable to the advection timescale.  Importantly, the magnitude of the asymmetry is also about an order-of-magnitude smaller than shown in \citet{Tamb14}.  In short, we find no evidence for LESA in our models, but they are limited to 2D and more work is needed before a final judgment on the existence of LESA can be made.  On the other hand, in our 2D simulations we do find that the dipolar component of the \textit{sum} $F_{\nu_e}+F_{\bar{\nu}_e}$ (and also the sum of number fluxes) is highly correlated with the dipolar component of the shock, as shown in the top panel of Fig.~\ref{fig:lesa}.  In this panel, $a_1/a_0$ of $F_{\nu_e}+F_{\bar{\nu}_e}$ is shown, multiplied by ten for ease of comparison with $a_1/a_0$ of the shock.

\subsection{Temporal Variations}
\label{temporal}
\begin{figure}
\includegraphics[width=0.8\textwidth]{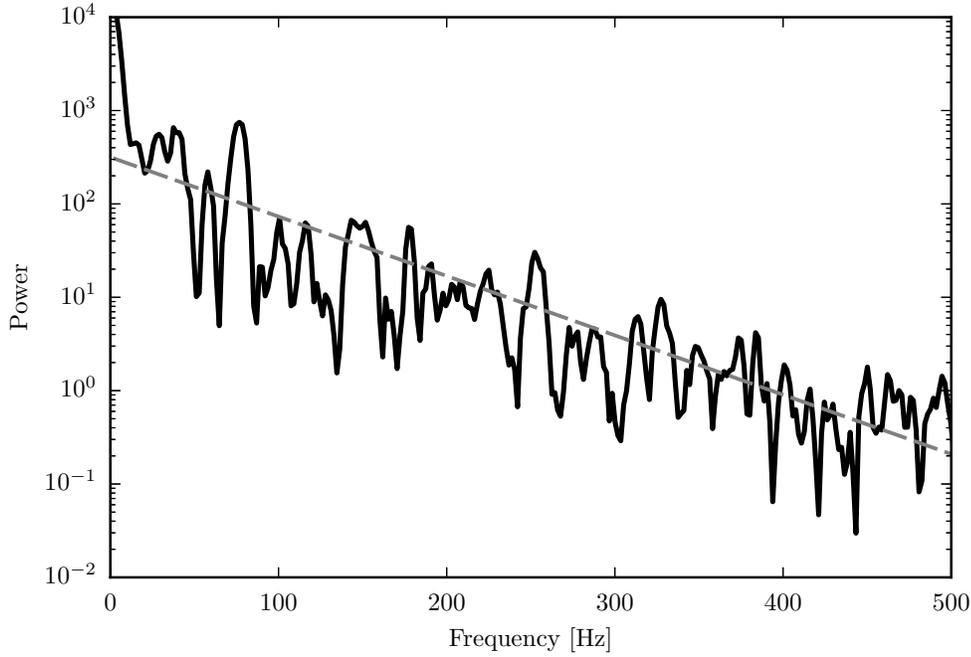}
\caption{Power spectrum of the simulated IceCube detection signal, as discussed in Sec.~\ref{temporal}.  The gray dashed line shows a best-fit exponential.  The normalization and slope of our power spectrum seems to agree quite well with the results of \citet{Lund10}, who had to compute hemispheric-averages of their ``ray-by-ray'' data to approximate the effects of multi-D transport.}
\label{fig:nu_pspec}
\end{figure}

An interesting diagnostic of the temporal variations in the neutrino sector is the power spectrum of the signal expected to be detected from a galactic supernova.  Following \citet{Lund10}, we compute the event rate expected for the IceCube detector at 256 observer orientations for our $25$-$\msun$ model, assumed to be at a distance of 10 kpc.  We compute the power spectra of these signals and then solid-angle average these to produce the power spectrum shown in Fig.~\ref{fig:nu_pspec}.  Comparing with Fig.~5 of \citet{Lund10}, we see remarkable agreement between our power spectrum and the ``north hemispheric average'' reported in that work.  Interestingly, their power spectrum produced by averaging the spectra from each ray, in a manner similar to what we have done, shows significantly more power, particularly at high frequencies.  Their hemispheric averaging is meant to mask the inherent error in their ``ray-by-ray'' transport by mimicking the angular integral character of transport.  Evidently, our MGFLD transport naturally produces power spectra that agree quite nicely with their hemisphere-averaged results.  By contrast, the ``ray-by-ray'' technique clearly leads to significant overestimates of the variability in the quantity $L_{\bar{\nu}_e} \varepsilon_{\rm rms}^2$, particularly at high frequencies.  This quantity bears directly on the energy deposition in the gain region, and so directly on the viability of the neutrino-driven supernova mechanism.

\begin{figure}
\includegraphics[width=0.8\textwidth]{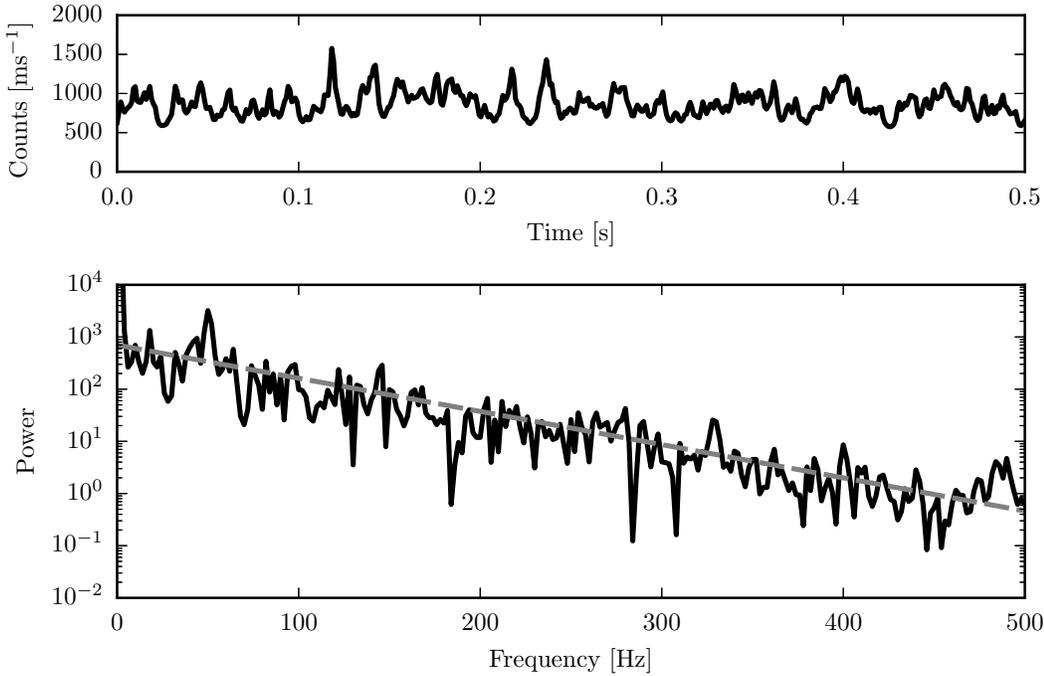}
\caption{The top panel shows a fake IceCube signal, produced as a sum of a constant and 500 randomly placed Lorentzian pulses.  The bottom panel shows the power spectrum of this fake signal, with the dashed gray line showing the expected exponential fall off for the pulse width adopted in the fake signal, $\exp(-4\pi f\gamma)$, where $\gamma$ is the half-width at half-maximum, as discussed in Sec.~\ref{temporal}.}
\label{fig:lorentz}
\end{figure}

One interesting aspect of the power spectrum that was not addressed by \citet{Lund10} (or the follow-on references \citealt{Lund12} and \citealt{Tamb13}) is its exponential shape.  The dashed line in Fig.~\ref{fig:nu_pspec} is a fit of the power spectrum above $50\,{\rm Hz}$ with $P(f)=P_0 \exp (-4\pi f\tau)$, where $P_0$ and $\tau$ are free parameters.  The resulting fit seems to describe our results quite well, and appears to be consistent with what others have found.  At least one way to produce an exponential power spectrum is with Lorentzian pulses of the form
\begin{equation}
L(t) \propto \frac{\gamma}{(t-t_0)^2 + \gamma^2}\;,
\end{equation}
where $t_0$ is the time of the pulse and $\gamma$ is the half-width at half-maximum.  In Fig.~\ref{fig:lorentz}, we show a fake signal, produced as a constant plus 500 Lorentzian pulses\footnote{The number of pulses only effects the normalization, not the e-folding timescale of the exponential.} at randomly chosen times and with randomly chosen amplitudes.  The power spectrum, shown below the signal, has an exponential dependence on frequency, with an e-folding timescale of $4\pi\gamma$.  So, one interpretation of the exponential power spectrum of our modeled IceCube signal is that the signal is comprised of many approximately Lorentzian pulses with a typical full-width at half-maximum timescale, extracted from our fit, of $2\tau=2.3\ms$.  This represents very rapid variability.  A simplistic picture of how this timescale may emerge is of blobs cooling as they advect through the cooling region.  In this picture, we adopt the simple scaling $L_{\rm blob} \varepsilon_{\rm rms}^2\sim T^8$ and estimate the timescale $2\tau\sim(8 v_r d\ln T/dr)^{-1}$, with $v_r$ the radial velocity.  Using the velocity and temperature gradient in the cooling region shown in Fig.~\ref{fig:rbr_comp}, our estimate yields $2\tau\sim2.5\ms$, in surprisingly good agreement with the measured value of $2.3\ms$.

\section{Summary and Conclusions}
\label{sum}

Using our new multi-group, multi-dimensional radiation hydrodynamics code CASTRO, which incorporates all terms to $\mathcal{O}(v/c)$ in the transport and does not make the ray-by-ray approximation employed by all other groups now modeling core-collapse supernovae, we have simulated in two spatial dimensions the dynamics of four progenitor massive star models. One goal was to determine, using a different code, whether the outcome of our previous simulations using the VULCAN/2D methodology \citep{Burr06,Burr07,Ott08} depended upon the absence of the $\mathcal{O}(v/c)$ terms in VULCAN/2D. We have determined that the results are qualitatively the same and, as when employing VULCAN/2D, we do not see explosions by the neutrino heating mechanism after $\sim$600 milliseconds after bounce. Both codes perform two-dimensional transport, though using a multi-group flux-limited (MGFLD) formulation.  This conclusion concerning the overall outcome of these models (i.e., explosion in 2D, driven by neutrino heating) is in contrast with the results of \citet{Brue13} and Janka et al. (Nuclear Astrophysics Workshop, Ringberg Castle, 2014), who also do not agree one with the other, but who do obtain neutrino-driven explosions in some or all of their 2D simulations.

One is left to ponder the reasons for these remaining differences in the community of researchers engaged in detailed simulations of the core-collapse supernova phenomenon. We have demonstrated that the ray-by-ray approach does not reproduce the correct angular and temporal neutrino field variations, though no one has yet performed the head-to-head ray-by-ray versus correct transport comparisons needed to definitively clarify the impact of the ray-by-ray approximation. We speculate, however, that the combination of the ray-by-ray approach with the artificiality of the axial sloshing effects manifest in 2D simulations may be the reason the groups using ray-by-ray obtain explosions in 2D (when they do).

While the ray-by-ray approximation is clearly suspect, there are other differences that may prove to play an important role in producing the range of findings in the community.  One might suspect that differences in the neutrino interaction physics may play an important role, but our experimentation indicates that the numerous hydrodynamic, thermal, and radiative feedbacks in the core-collapse problem mute the effects of even large changes in the neutrino-matter cross sections and associated emissivities on the dynamic evolution after collapse.  In 1D test calculations we have performed in which the $\nu_{e}$$-$neutron absorption cross section was changed by a factor of two (both increased and decreased), the resulting stalled shock radii were the same to within a few percent.  Some recent calculations suggest there may be some sensitivity to the choice of equation of state (EOS), with calculations using the Lattimer and Swesty EOS tending to explode more easily than those using the Shen EOS \citep{Janka12,Suwa13,Couch13}.  Since both the present study and the VULCAN/2D studies used the Shen EOS and failed to explode, it may prove illuminating to repeat some of these calculations with the Lattimer and Swesty EOS.  The effects of general relativity (GR) and the differing fidelity with which they are included in calculations may also contribute \citep[e.g.][]{Mull12}, but note that GR seems not to be generally requisite for explosions, as demonstrated by the 2D $27$-$\msun$ models reported in \citet{Mull12_sasi} that included GR and in \citet{Hank13} that used a monopolar gravity approximation with mock GR corrections, nevertheless transitioning to explosion at nearly the same post-bounce time.  The marked difference between 1D and 2D in the early evolution of the shock radius in our models, which we attribute to a vigorous burst of prompt convection seeded by perturbations from our aspherical grid, may also be a concern, but we would expect the memory of this defect to be lost within a few dynamical times ($<100\ms$) as the system dynamically relaxes to a quasi-steady configuration.  Differences in the transport algorithms (apart from the ray-by-ray versus multi-D transport issue) could be to blame, and code-to-code comparisons are called for.  This was one early motivation for embarking upon this study with CASTRO---to see whether the outcomes were different from those we obtained using VULCAN/2D. But, more inter-group comparisons, not just intra-group comparisons, are needed.

The fact that the 3D simulations of the Garching group are not exploding when they were in 2D \citep{Hank13} should be a wake-up call to the community to determine the origins of these differences between the various simulation groups and between 2D and 3D.  As we have suggested, the use of the ray-by-ray approach is dubious, and since its artificial character is more manifest in 2D we suspect that it is part of the problem. However, this does not explain the current conundrum in 3D---something else may be amiss. It could be that the progenitor models are to blame and a new generation of such models, performed in 3D at the terminal stages of a massive star's life are needed \citep{Meak11}. It could be that rotation, even the modest rotation expected from the pulsar injection constraint \citep{Emme89}, could, by the resultant centrifugal support and consequent expected boost in the stalled shock radius, convert duds into explosions.  This is the simplest solution, and one is reminded that the exploding model of \citet{Mare09} was rotating. Both large-scale and turbulent magnetic fields could play a role, through the associated stress, but also due to enhanced angular momentum transport from the core to the mantle \citep[e.g.][]{Sawa14}.  However, without very rapid rotation, which might be associated with the rare hypernovae \citep{Burr07mhd} that serve as a bridge to the collapsar model of long-soft gamma-ray bursts, there would not seem to be enough extra free energy to power explosion generically.  Perturbations of the progenitor cores that collapse have never been properly included into supernova theory, and might be a fruitful line of investigation \citep{Couc13}. Such perturbations seed the instabilities long identified with more robust dynamics and the viability of the delayed neutrino mechanism.

Whatever the solution to this recalcitrant problem, advances in the numerical arts seem destined to play a central role.  Approximations have been made by all groups to accommodate the limitations of the available computer resources, leaving one to wonder whether such compromises have corrupted the results.  One would hope that simple, compelling reasoning, and physical insight could in the end lead to a solution.  This has happened before in astrophysics. However, the complexity of the dynamics, the fact that the explosion energy is a small fraction of the available energy, and the circumstance that the central ``engine'' is shrouded in mystery by the profound opacity of the stellar envelope, and, hence, is itself (almost) inaccessible to direct observation or measurement, may mitigate against a breakthrough unaided by computation.

\acknowledgments

The authors acknowledge conversations and collaborations with Jeremiah Murphy, 
Christian Ott, Stan Woosley, Ann Almgren, John Bell, and Louis Howell.
The development of the CASTRO code was supported by the Scientific Discovery through
Advanced Computing (SciDAC) program of the DOE, under grant number DE-FG02-08ER41544,
the NSF under the subaward no. ND201387 to the Joint Institute for Nuclear Astrophysics (JINA, NSF PHY-0822648),
and the NSF PetaApps program, under award OCI-0905046 via a subaward
no. 44592 from Louisiana State University to Princeton University.
The authors employed computational resources provided by the TIGRESS
high performance computer center at Princeton University, which is jointly supported by the Princeton
Institute for Computational Science and Engineering (PICSciE) and the Princeton University Office of
Information Technology; by the National Energy Research Scientific Computing Center
(NERSC), which is supported by the Office of Science of the US Department of
Energy under contract DE-AC03-76SF00098; and on the Kraken supercomputer,
hosted at NICS and provided by the National Science Foundation through
the TeraGrid Advanced Support Program under grant number TG-AST100001.

\appendix

\section{Approximate Multi-angle and ``Ray-by-ray'' Transport}\label{appendix}
For comparisons between the various approaches to neutrino transport, we have computed fluxes based on approximate multi-angle and ``ray-by-ray'' calculations of the specific intensities.  We make three simplifying assumptions: 1) the problem is treated as time-independent, 2) we drop all $\mathcal{O}(v/c)$ terms, and 3) we ignore all scattering processes.  Under these assumptions, the transport equation along any ray reduces to
\begin{equation}\label{eq:transport}
\frac{dI_\varepsilon}{ds} = j_\varepsilon - c\kappa_\varepsilon I_\varepsilon\,,
\end{equation}
where $I_\varepsilon$ is the specific intensity, $s$ is the path length along the ray, $j_\varepsilon$ is the volume emissivity, and $\kappa_\varepsilon$ is the absorption cross section, all defined at neutrino energy $\varepsilon$.

In order to compute the fluxes to facilitate a comparison of multi-angle and ``ray-by-ray'' with our MGFLD results, we must compute the direction-dependent specific intensities at each point of comparison so that the fluxes can be computed by taking the first moment.  We proceed as follows.  First, at each point of comparison, we setup a local spherical coordinate system ($\tilde{r}$, $\tilde{\theta}$, $\tilde{\phi}$) with the $\tilde{\theta}=0$ axis oriented along the local radial direction of the simulation.  We then compute the intensities $I(\tilde{\theta},\tilde{\phi})$ at discrete angles by solving Eq.\ref{eq:transport} along the ray passing through our point at angles $\tilde{\theta}$ and $\tilde{\phi}$ to the radial direction.  

For the multi-angle calculation, we compute intensities between $\tilde{\theta}=0$ and $\tilde{\theta}=\tilde{\theta}_{\rm max}$, where $\tilde{\theta}_{\rm max}$ is the maximum angular extent of the shock, within which essentially all of the emission/absorption occurs.  At each $\tilde{\theta}$, we must also compute intensities as a function of $\tilde{\phi}$, but the axisymmetry of the underlying simulation implies $I(\tilde{\theta},\tilde{\phi})=I(\tilde{\theta},-\tilde{\phi})$, which allows us to consider only $\tilde{\phi}$ in the range $[0,\pi]$.

For the ``ray-by-ray'' calculation, we must only compute $I(\tilde{\theta})$ since the underlying assumption of a spherical distribution of matter implies that the intensities are symmetric about the local radial direction.  This is, of course, the main motivation behind the ``ray-by-ray'' scheme---there are far fewer intensities to compute and they depend on far less of the simulation data.

\newpage
\bibliographystyle{apj}
\bibliography{refs}

\begin{thebibliography}{54}
\expandafter\ifx\csname natexlab\endcsname\relax\def\natexlab#1{#1}\fi

\bibitem[{{Almgren} {et~al.}(2010){Almgren}, {Beckner}, {Bell}, {Day},
  {Howell}, {Joggerst}, {Lijewski}, {Nonaka}, {Singer}, \& {Zingale}}]{Almg10}
{Almgren}, A.~S., {et~al.} 2010, \apj, 715, 1221

\bibitem[{{Brandt} {et~al.}(2011){Brandt}, {Burrows}, {Ott}, \&
  {Livne}}]{Bran11}
{Brandt}, T.~D., {Burrows}, A., {Ott}, C.~D., \& {Livne}, E. 2011, \apj, 728, 8

\bibitem[{{Bruenn} {et~al.}(2013){Bruenn}, {Mezzacappa}, {Hix}, {Lentz},
  {Bronson Messer}, {Lingerfelt}, {Blondin}, {Endeve}, {Marronetti}, \&
  {Yakunin}}]{Brue13}
{Bruenn}, S.~W., {et~al.} 2013, \apjl, 767, L6

\bibitem[{{Buras} {et~al.}(2006{\natexlab{a}}){Buras}, {Janka}, {Rampp}, \&
  {Kifonidis}}]{Bura06}
{Buras}, R., {Janka}, H.-T., {Rampp}, M., \& {Kifonidis}, K.
  2006{\natexlab{a}}, \aap, 457, 281

\bibitem[{{Buras} {et~al.}(2006{\natexlab{b}}){Buras}, {Rampp}, {Janka}, \&
  {Kifonidis}}]{Bura06a}
{Buras}, R., {Rampp}, M., {Janka}, H.-T., \& {Kifonidis}, K.
  2006{\natexlab{b}}, \aap, 447, 1049

\bibitem[{{Burrows}(2013)}]{Burr13}
{Burrows}, A. 2013, Reviews of Modern Physics, 85, 245

\bibitem[{{Burrows} {et~al.}(2007{\natexlab{a}}){Burrows}, {Dessart}, {Livne},
  {Ott}, \& {Murphy}}]{Burr07}
{Burrows}, A., {Dessart}, L., {Livne}, E., {Ott}, C.~D., \& {Murphy}, J.
  2007{\natexlab{a}}, \apj, 664, 416

\bibitem[{{Burrows} {et~al.}(2007{\natexlab{b}}){Burrows}, {Dessart}, {Livne},
  {Ott}, \& {Murphy}}]{Burr07mhd}
---. 2007{\natexlab{b}}, \apj, 664, 416

\bibitem[{{Burrows} {et~al.}(2012){Burrows}, {Dolence}, \& {Murphy}}]{Burr12}
{Burrows}, A., {Dolence}, J.~C., \& {Murphy}, J.~W. 2012, \apj, 759, 5

\bibitem[{{Burrows} \& {Goshy}(1993)}]{Burr93}
{Burrows}, A., \& {Goshy}, J. 1993, \apjl, 416, L75

\bibitem[{{Burrows} {et~al.}(1995){Burrows}, {Hayes}, \& {Fryxell}}]{Burr95}
{Burrows}, A., {Hayes}, J., \& {Fryxell}, B.~A. 1995, \apj, 450, 830

\bibitem[{{Burrows} {et~al.}(2006){Burrows}, {Livne}, {Dessart}, {Ott}, \&
  {Murphy}}]{Burr06}
{Burrows}, A., {Livne}, E., {Dessart}, L., {Ott}, C.~D., \& {Murphy}, J. 2006,
  \apj, 640, 878

\bibitem[{{Castor}(2004)}]{Cast04}
{Castor}, J.~I. 2004, {Radiation Hydrodynamics}

\bibitem[{{Colgate} \& {White}(1966)}]{Colg66}
{Colgate}, S.~A., \& {White}, R.~H. 1966, \apj, 143, 626

\bibitem[{{Couch}(2013{\natexlab{a}})}]{Couc13_2d3d}
{Couch}, S.~M. 2013{\natexlab{a}}, \apj, 775, 35

\bibitem[{{Couch}(2013{\natexlab{b}})}]{Couch13}
---. 2013{\natexlab{b}}, \apj, 765, 29

\bibitem[{{Couch} \& {Ott}(2013)}]{Couc13}
{Couch}, S.~M., \& {Ott}, C.~D. 2013, \apjl, 778, L7

\bibitem[{{Dessart} {et~al.}(2006){Dessart}, {Burrows}, {Livne}, \&
  {Ott}}]{Dess06}
{Dessart}, L., {Burrows}, A., {Livne}, E., \& {Ott}, C.~D. 2006, \apj, 645, 534

\bibitem[{{Dolence} {et~al.}(2013){Dolence}, {Burrows}, {Murphy}, \&
  {Nordhaus}}]{Dole13}
{Dolence}, J.~C., {Burrows}, A., {Murphy}, J.~W., \& {Nordhaus}, J. 2013, \apj,
  765, 110

\bibitem[{{Emmering} \& {Chevalier}(1989)}]{Emme89}
{Emmering}, R.~T., \& {Chevalier}, R.~A. 1989, \apj, 345, 931

\bibitem[{Falgout \& Yang(2002)}]{Falg02}
Falgout, R.~D., \& Yang, U.~M. 2002, in Computational Science---ICCS 2002
  (Springer Berlin Heidelberg), 632--641

\bibitem[{{Hanke} {et~al.}(2013){Hanke}, {M{\"u}ller}, {Wongwathanarat},
  {Marek}, \& {Janka}}]{Hank13}
{Hanke}, F., {M{\"u}ller}, B., {Wongwathanarat}, A., {Marek}, A., \& {Janka},
  H.-T. 2013, \apj, 770, 66

\bibitem[{{Herant} {et~al.}(1992){Herant}, {Benz}, \& {Colgate}}]{Hera92}
{Herant}, M., {Benz}, W., \& {Colgate}, S. 1992, \apj, 395, 642

\bibitem[{{Hubeny} \& {Burrows}(2007)}]{Hube07}
{Hubeny}, I., \& {Burrows}, A. 2007, \apj, 659, 1458

\bibitem[{{Janka}(2012)}]{Janka12}
{Janka}, H.-T. 2012, Annual Review of Nuclear and Particle Science, 62, 407

\bibitem[{{Janka} \& {Mueller}(1996)}]{Jank96}
{Janka}, H.-T., \& {Mueller}, E. 1996, \aap, 306, 167

\bibitem[{{Liebend{\"o}rfer} {et~al.}(2009){Liebend{\"o}rfer}, {Whitehouse}, \&
  {Fischer}}]{Lieb09}
{Liebend{\"o}rfer}, M., {Whitehouse}, S.~C., \& {Fischer}, T. 2009, \apj, 698,
  1174

\bibitem[{{Lund} {et~al.}(2010){Lund}, {Marek}, {Lunardini}, {Janka}, \&
  {Raffelt}}]{Lund10}
{Lund}, T., {Marek}, A., {Lunardini}, C., {Janka}, H.-T., \& {Raffelt}, G.
  2010, \prd, 82, 063007

\bibitem[{{Lund} {et~al.}(2012){Lund}, {Wongwathanarat}, {Janka}, {M{\"u}ller},
  \& {Raffelt}}]{Lund12}
{Lund}, T., {Wongwathanarat}, A., {Janka}, H.-T., {M{\"u}ller}, E., \&
  {Raffelt}, G. 2012, \prd, 86, 105031

\bibitem[{{Marek} \& {Janka}(2009)}]{Mare09}
{Marek}, A., \& {Janka}, H.-T. 2009, \apj, 694, 664

\bibitem[{{Meakin} {et~al.}(2011){Meakin}, {Sukhbold}, \& {Arnett}}]{Meak11}
{Meakin}, C.~A., {Sukhbold}, T., \& {Arnett}, W.~D. 2011, \apss, 336, 123

\bibitem[{{Mezzacappa} {et~al.}(2014){Mezzacappa}, {Bruenn}, {Lentz}, {Hix},
  {Messer}, {Harris}, {Lingerfelt}, {Endeve}, {Yakunin}, {Blondin}, \&
  {Marronetti}}]{Mezz14}
{Mezzacappa}, A., {et~al.} 2014, in Astronomical Society of the Pacific
  Conference Series, Vol. 488, 8th International Conference of Numerical
  Modeling of Space Plasma Flows (ASTRONUM 2013), ed. N.~V. {Pogorelov},
  E.~{Audit}, \& G.~P. {Zank}, 102

\bibitem[{{Miller} \& {Colella}(2002)}]{Mill02}
{Miller}, G.~H., \& {Colella}, P. 2002, Journal of Computational Physics, 183,
  26

\bibitem[{{Mueller} \& {Janka}(2014)}]{Mull14}
{Mueller}, B., \& {Janka}, H.-T. 2014, ArXiv e-prints

\bibitem[{{M{\"u}ller} {et~al.}(2010){M{\"u}ller}, {Janka}, \&
  {Dimmelmeier}}]{Mull10}
{M{\"u}ller}, B., {Janka}, H.-T., \& {Dimmelmeier}, H. 2010, \apjs, 189, 104

\bibitem[{{M{\"u}ller} {et~al.}(2012{\natexlab{a}}){M{\"u}ller}, {Janka}, \&
  {Heger}}]{Mull12_sasi}
{M{\"u}ller}, B., {Janka}, H.-T., \& {Heger}, A. 2012{\natexlab{a}}, \apj, 761,
  72

\bibitem[{{M{\"u}ller} {et~al.}(2012{\natexlab{b}}){M{\"u}ller}, {Janka}, \&
  {Marek}}]{Mull12}
{M{\"u}ller}, B., {Janka}, H.-T., \& {Marek}, A. 2012{\natexlab{b}}, \apj, 756,
  84

\bibitem[{{Murphy} \& {Burrows}(2008)}]{Murp08}
{Murphy}, J.~W., \& {Burrows}, A. 2008, \apj, 688, 1159

\bibitem[{{Murphy} {et~al.}(2013){Murphy}, {Dolence}, \& {Burrows}}]{Murp13}
{Murphy}, J.~W., {Dolence}, J.~C., \& {Burrows}, A. 2013, \apj, 771, 52

\bibitem[{{Ott} {et~al.}(2008){Ott}, {Burrows}, {Dessart}, \& {Livne}}]{Ott08}
{Ott}, C.~D., {Burrows}, A., {Dessart}, L., \& {Livne}, E. 2008, \apj, 685,
  1069

\bibitem[{{Pejcha} \& {Thompson}(2012)}]{Pejc12}
{Pejcha}, O., \& {Thompson}, T.~A. 2012, \apj, 746, 106

\bibitem[{{Sawai} \& {Yamada}(2014)}]{Sawa14}
{Sawai}, H., \& {Yamada}, S. 2014, ArXiv e-prints

\bibitem[{{Shen} {et~al.}(1998{\natexlab{a}}){Shen}, {Toki}, {Oyamatsu}, \&
  {Sumiyoshi}}]{Shen98a}
{Shen}, H., {Toki}, H., {Oyamatsu}, K., \& {Sumiyoshi}, K. 1998{\natexlab{a}},
  Nuclear Physics A, 637, 435

\bibitem[{{Shen} {et~al.}(1998{\natexlab{b}}){Shen}, {Toki}, {Oyamatsu}, \&
  {Sumiyoshi}}]{Shen98b}
---. 1998{\natexlab{b}}, Progress of Theoretical Physics, 100, 1013

\bibitem[{{Sumiyoshi} {et~al.}(2014){Sumiyoshi}, {Takiwaki}, {Matsufuru}, \&
  {Yamada}}]{Sumi14}
{Sumiyoshi}, K., {Takiwaki}, T., {Matsufuru}, H., \& {Yamada}, S. 2014, ArXiv
  e-prints

\bibitem[{{Suwa} {et~al.}(2010){Suwa}, {Kotake}, {Takiwaki}, {Whitehouse},
  {Liebend{\"o}rfer}, \& {Sato}}]{Suwa10}
{Suwa}, Y., {Kotake}, K., {Takiwaki}, T., {Whitehouse}, S.~C.,
  {Liebend{\"o}rfer}, M., \& {Sato}, K. 2010, \pasj, 62, L49

\bibitem[{{Suwa} {et~al.}(2013){Suwa}, {Takiwaki}, {Kotake}, {Fischer},
  {Liebend{\"o}rfer}, \& {Sato}}]{Suwa13}
{Suwa}, Y., {Takiwaki}, T., {Kotake}, K., {Fischer}, T., {Liebend{\"o}rfer},
  M., \& {Sato}, K. 2013, \apj, 764, 99

\bibitem[{{Takiwaki} {et~al.}(2012){Takiwaki}, {Kotake}, \& {Suwa}}]{Taki12}
{Takiwaki}, T., {Kotake}, K., \& {Suwa}, Y. 2012, \apj, 749, 98

\bibitem[{{Takiwaki} {et~al.}(2013){Takiwaki}, {Kotake}, \& {Suwa}}]{Taki13}
---. 2013, ArXiv e-prints

\bibitem[{{Tamborra} {et~al.}(2014){Tamborra}, {Hanke}, {Janka}, {Mueller},
  {Raffelt}, \& {Marek}}]{Tamb14}
{Tamborra}, I., {Hanke}, F., {Janka}, H.-T., {Mueller}, B., {Raffelt}, G.~G.,
  \& {Marek}, A. 2014, ArXiv e-prints

\bibitem[{{Tamborra} {et~al.}(2013){Tamborra}, {Hanke}, {M{\"u}ller}, {Janka},
  \& {Raffelt}}]{Tamb13}
{Tamborra}, I., {Hanke}, F., {M{\"u}ller}, B., {Janka}, H.-T., \& {Raffelt}, G.
  2013, Physical Review Letters, 111, 121104

\bibitem[{{Woosley} \& {Heger}(2007)}]{Woos07}
{Woosley}, S.~E., \& {Heger}, A. 2007, \physrep, 442, 269

\bibitem[{{Zhang} {et~al.}(2011){Zhang}, {Howell}, {Almgren}, {Burrows}, \&
  {Bell}}]{Zhan11}
{Zhang}, W., {Howell}, L., {Almgren}, A., {Burrows}, A., \& {Bell}, J. 2011,
  \apjs, 196, 20

\bibitem[{{Zhang} {et~al.}(2013){Zhang}, {Howell}, {Almgren}, {Burrows},
  {Dolence}, \& {Bell}}]{Zhan13}
{Zhang}, W., {Howell}, L., {Almgren}, A., {Burrows}, A., {Dolence}, J., \&
  {Bell}, J. 2013, \apjs, 204, 7

\end{thebibliography}

\end{document}